%% file: Highz_clustering_final.tex
\documentclass[apj, numberedappendix]{emulateapj}
\usepackage{graphicx, subfigure}
\usepackage{amsmath, natbib, listings}
\usepackage[bottom, hang, flushmargin]{footmisc}
\usepackage[urlcolor=magenta,citecolor=green,linkcolor=red]{hyperref}
\usepackage{appendix}
\usepackage{rotate}
\usepackage{adjustbox}
\usepackage[normalem]{ulem}

\begin{document}

\shorttitle{High-$z$ Quasar Clustering}
\shortauthors{}

\title{The Clustering of High-Redshift ($2.9 \leq \MakeLowercase{z} \leq 5.1
$) Quasars in SDSS Stripe 82}

\author{
  John D. Timlin\altaffilmark{1,$\star$},
  Nicholas P. Ross\altaffilmark{2}, 
 Gordon T. Richards\altaffilmark{1},
Adam D. Myers\altaffilmark{3},
Andrew Pellegrino\altaffilmark{1},
Franz E. Bauer\altaffilmark{4,5,6},
Mark Lacy\altaffilmark{7},
Donald P. Schneider\altaffilmark{8,9},
Edward J. Wollack\altaffilmark{10},
Nadia L. Zakamska\altaffilmark{11}
}

\altaffiltext{$\star$}{\email{john.d.timlin@drexel.edu}}
\altaffiltext{1}{Department of Physics, Drexel University, 3141 Chestnut Street, Philadelphia, PA 19104, U.S.A}
\altaffiltext{2}{Institute for Astronomy, University of Edinburgh, Royal Observatory, Edinburgh, EH9 3HJ, U.K.}
\altaffiltext{3}{Department of Physics and Astronomy, University of Wyoming, 1000 University Ave., Laramie, WY, 82071, USA}

\altaffiltext{4}{Instituto de Astrof{\'{\i}}sica and Centro de Astroingenier{\'{\i}}a, Facultad de F{\'{i}}sica, Pontificia Universidad Cat{\'{o}}lica de Chile, Casilla 306, Santiago 22, Chile} 
\altaffiltext{5}{Millennium Institute of Astrophysics (MAS), Nuncio Monse{\~{n}}or S{\'{o}}tero Sanz 100, Providencia, Santiago, Chile} 
\altaffiltext{6}{Space Science Institute, 4750 Walnut Street, Suite 205, Boulder, Colorado 80301} 

\altaffiltext{7}{National Radio Astronomy Observatory, 520 Edgemont Road, Charlottesville, VA 22903, U.S.A}

\altaffiltext{8}{Department of Astronomy \& Astrophysics, 525 Davey Lab, The Pennsylvania State University, University Park, PA 16802, USA}
\altaffiltext{9}{Institute for Gravitation and the Cosmos, The Pennsylvania State University, University Park, PA 16802, USA}

\altaffiltext{10}{NASA Goddard Space Flight Center, Greenbelt, MD 20771}
\altaffiltext{11}{Department of Physics and Astronomy, Johns Hopkins University, Bloomberg Center, 3400 N. Charles St., Baltimore, MD 21218, USA}

\date{\today}

\begin{abstract}
We present a measurement of the two-point autocorrelation function of photometrically-selected, high-$z$ quasars over $\sim$ 100 deg$^2$ on the Sloan Digitial Sky Survey Stripe 82 field. Selection is performed using three machine-learning algorithms in a six-dimensional, optical/mid-infrared color space. Optical data from the Sloan Digitial Sky Survey is combined with overlapping deep mid-infrared data from the \emph{Spitzer} IRAC Equatorial Survey and the \emph{Spitzer}-HETDEX Exploratory Large-area survey. Our selection algorithms are trained on the colors of known high-$z$ quasars. The selected quasar sample consists of 1378 objects and contains both spectroscopically-confirmed quasars and photometrically-selected quasar candidates. These objects span a redshift range of $2.9 \leq z \leq 5.1$ and are generally fainter than $i=20.2$; a regime which has lacked sufficient number density to perform autocorrelation function measurements of photometrically-classified quasars. We compute the angular correlation function of these data, marginally detecting quasar clustering. We fit a single power-law with an index of $\delta = 1.39 \pm 0.618$ and amplitude of $\theta_0 = 0.71 \pm 0.546$ arcmin. A dark-matter model is fit to the angular correlation function to estimate the linear bias. At the average redshift of our survey ($\langle z \rangle = 3.38$) the bias is $b = 6.78 \pm 1.79$. Using this bias, we calculate a characteristic dark-matter halo mass of 1.70--9.83$\times 10^{12}h^{-1} M_{\odot}$. Our bias estimate suggests that quasar feedback intermittently shuts down the accretion of gas onto the central super-massive black hole at early times. If confirmed, these results hint at a level of luminosity dependence in the clustering of quasars at high-$z$.
\end{abstract}

\keywords{quasars: High-Redshift; Clustering; machine-learning}

\maketitle

\section{Introduction}\label{sec:1}
In the present day Universe, super-massive black holes (SMBHs) reside at the center of most, if not all, galaxies with $M_{\star} \gtrsim 10^{10} M_{\odot}$, in which star-formation has almost completely ceased \citep[e.g.,][]{Bell2008, Bower2017}. It is commonly accepted that every massive galaxy has undergone {\it at least} one quasar phase within its lifetime (\citealt{Soltan1982}; \citealt{Richstone1998}). In this quasar phase, baryons in an accretion disk lose angular momentum through mechanisms such as viscous transfer, and eventually are accreted by the SMBH (\citealt{Salpeter1964}, \citealt{Lynden-Bell1969}, \citealt{Rees1984}). The friction in the disk heats the baryons causing the disk to shine in the optical, ultraviolet (UV), and X-rays. 

Quasars, defined here as a luminous active galactic nuclei with bolometric luminosity $L_{\rm bol}$ above $\sim 10^{45}$ erg s$^{-1}$, are among the most luminous objects in the Universe, and therefore, can trace the large scale structure out to high redshift. Galaxies are thought to reside in the peaks in the dark-matter (DM) distribution, and are generally biased tracers of the underlying DM \citep[e.g., ][]{DekelLahav1999, ShethTormen1999, Peacock_book1999}. This relationship can be quantified by measuring the linear bias parameter, $b$. As an initial guide, we define $b$ as: 
\begin{equation}\label{bias}
\delta_{\rm Q} = b \; \delta_{\rm DM}
\end{equation}
where $\delta_{\rm Q}$ is the quasar density contrast and $\delta_{\rm DM}$ is the mass density contrast. Defining the two-point auto-correlation function (2PCF) as $\xi(r) = \left \langle \delta(x) \delta(x + r) \right \rangle$, where $r$ is the separation between two local over-densities, leads to 
\begin{equation}
\xi_{\rm Q}(r) = b^{2}_{\rm Q} \; \xi_{\rm DM}(r)
\end{equation} 
where $\xi_{\rm Q}$ is the quasar two-point correlation function and $\xi_{\rm DM}$ is the DM correlation function. The 2PCF is defined as the joint probability of finding a pair of objects having a particular separation in two volume elements (\citealt{Totsuji1969}; \citealt{Peebles1980}) and is a statistic commonly employed to measure the spatial distribution of galaxies \citep[e.g.,][]{Zehavi2011}, hydrogen gas in absorption \citep[e.g.,][]{Bautista2017} and, in this case, quasars. In practice, the 2PCF is calculated as the excess probability, above a random Poisson distribution, of finding a pair of objects within an annulus between $r$ and $r+\delta r$ \citep{Peebles1980, MartinezSaar2002, Feigelson2012}.

\begin{deluxetable*}{l rr rccc l}
\tablecolumns{8}
\tablewidth{0pt}
\tablecaption{Selected quasar clustering measurements.}
\tablehead{
\colhead{Survey}  &  \colhead{Area}         & \colhead{N$_{\rm Q}$} & \colhead{Magnitude} & \colhead{Selection}       & \colhead{$z$-range} & \colhead{Type$^{a}$}   &	\colhead{Reference} \\
  \colhead{}  &  \colhead{/ deg$^{2}$}  &   \colhead{} &  \colhead{range} &\colhead{} &\colhead{}&\colhead{}&\colhead{}
}
\startdata
\input{tab1}\label{tab:other_clustering}
\enddata
\tablecomments{$^{a}$Measurement of the auto-correlation function(A), the cross-correlation function(C), using  photometric(p)/spectroscopic(s) redshifts, or a combination of both (b). The studies in this table take advantage of the properties of quasars in X-ray (X), radio (R), mid-infrared (MIR), near-infrared (NIR), and optical (opt) wavelengths and use color-boxes (cb) and/or with machine-learning techniques for selection.   \\
$^{b}$``Extreme Deconvolution'', see \citet{Bovy2011}. \\
$^{c}$\citet{DiPompeo2014,DiPompeo2015} performed similar analyses on earlier WISE datasets. \\
$^{d}$Kernel Density Estimator, see \citet{Richards2009}. \\
$^{e}$\citet{Myers2006} performed a similar analysis on SDSS DR1 data. \\
}
\end{deluxetable*}

The 2PCF, and the corresponding bias, have been measured for quasars as a function of different observable properties, including redshift, luminosity and color;  Table~\ref{tab:other_clustering} presents a summary of recent results. Studies of quasar clustering as a function of luminosity \citep{daAngela2008, Shen2009, Eft2015, Chehade2016} have shown that the bias is very weakly, if at all, dependent on absolute quasar UV/optical luminosity. In fact, both \citet{Shen2013} and \citet{Krolewski2015} found no luminosity dependence of quasar clustering at low-$z$ by studying the cross-correlation between galaxies and quasars. This result implies that quasars all live in the most massive dark-matter halos, regardless of how bright the quasar shines. \citet{Aird2018}, however, suggested that the observed lack of luminosity dependence on quasar clustering may be due to a selection effect depending on the type of galaxy in which the AGN resides (star-forming or quiescent).

\citet{Croom2005}, \citet{Myers2007}, and \citet{Ross2009} have demonstrated that the bias evolves with redshift, increasing at higher redshift until the peak of quasar activity at $z\sim2.5$. These studies were performed with large number densities of either spectroscopically-confirmed or photometrically-selected quasars, driving down Poisson noise in the clustering measurement (see Table \ref{tab:other_clustering}). Interestingly, however, due to the evolution of the underlying DM density field, the masses of the halos quasars inhabit remains approximately constant at $M_{\rm halo}\sim 2-3 \times 10^{12} h^{-1} M_{\odot}$ from redshifts $z\sim2.5$ to the present day. \citet{Shen2007} performed a similar analysis of the luminous, high-$z$ ($2.9 \leq z \leq 5.4$) confirmed quasars from the Sloan Digital Sky Survey (SDSS; \citealt{York2000}) Data Release 5. Despite having low number densities ($\sim$ 1 quasar deg$^{-2}$), their study detected a large clustering signal, which implied that the bias increases rapidly beyond $z\sim2.5$, yielding a large increase in the DM halo mass estimate with redshift. 

Clustering has also been studied as function of quasar color, which is a proxy for quasar type. Here the results are not so definitive. \citet{Hickox2011} measured the clustering of both obscured and unobscured quasars, as defined by an optical-to-IR flux ratio (specifically $R_{AB}-[4.5]_{\rm Vega}$=6.0; \citealt{Hickox2007}), with bluer objects being classed as unobscured quasars. \citet{Hickox2011} reported ``marginally stronger clustering'' for the obscured quasars compared to the unobscured population, with the consequence that dust-obscured quasars tend to reside in more massive DM halos than `dust-free' quasars. \citet{Donoso2014}, using a similar selection to \citet{Hickox2011}, similarly found that obscured AGNs inhabit denser environments than unobscured AGNs. \citet{DiPompeo2014,DiPompeo2015,DiPompeo2016}, in finding a less significant difference between the clustering of obscured and unobscured quasars, noted that \citet{Donoso2014} discounted several critical systematics that affect the amplitude of quasar clustering measurements. 

Linking the measurements of the 2PCF and of the corresponding bias to quasar and host galaxy physical parameters is paramount in understanding the relationship between the observable Universe and the underlying DM distribution. These observables can then be used to direct theories and models of galaxy and quasar formation and evolution. One model that links the DM distribution, quasar activity and the associated environment was presented in \citet{Hopkins2007}. The simulations in that investigation predicted the clustering of the quasar population through the implementation of three different quasar feedback models. Quasar feedback works against gravity by forcing material away from the SMBH through radiation pressure, thus limiting the material that can accrete onto, and increase the mass of, the SMBH. This process can ultimately shut down the quasar phase and cause the SMBH to cease growing. Measuring the spatial distribution of quasars, particularly in the early Universe, can test the predictions made by \citet{Hopkins2007}. 

Testing these models requires surveys to push beyond the redshift peak in the quasar epoch ($2 \leq z \leq 3$; \citealt{Schmidt1995}; \citealt{Boyle2000}), \emph{and} delve further down the quasar luminosity function (QLF). Current surveys are underway to address this question, for example the extended Baryon Oscillation Spectroscopic Survey (eBOSS; \citealt{Dawson2016}), which will be able to select quasars out to $z \sim 3.5$ \citep{Myers2015}; however, the majority of existing quasar surveys are either designed to observe rest-frame UV bright quasars and/or are focused on $z<2$, thus new data and analysis is needed. 

In this paper we present the first measurements of the autocorrelation function of optical+infrared-selected quasars at $z>2.9$ with the 2PCF. This approach is made possible by the combination of deep optical data from the SDSS Stripe 82 coadded catalog (\citealt{Annis2014}; \citealt{Jiang2014}) as well as new deep, overlapping \emph{Spitzer} coverage from the \emph{Spitzer} IRAC Equatorial Survey (SpIES; \citealt{Timlin2016}) and the \emph{Spitzer}-HETDEX Exploratory Large-area (SHELA; \citealt{Papovich2016}) survey. Following the work of \citet{Richards2015}, we combine the color information from the optical and mid-infrared (MIR) and employ machine-learning algorithms to classify faint, high-$z$ quasar candidates using their photometric colors.

Traditionally, large numbers of quasars have been detected from \emph{Spitzer} surveys alone. Quasars tend to lie in a specific location in MIR color space so selection can be performed through various color cuts (\citealt{Lacy2004}; \citealt{Stern2005}). These constraints, while effective, lead to an increasing amount of contamination, particularly at high-$z$ where quasar colors overlap with the stellar locus, resulting in a higher level of incompleteness in the selection (\citealt{Assef2010}; \citealt{Donley2012}). \citet{Donley2012} added a power law selection requirement for classification, which significantly reduced contamination; however, quasar spectra are not necessarily power laws in the MIR \citep{Richards2015}. Similarly, optical-only selections have found a large number of new quasars in SDSS alone; however, these techniques suffer from incompleteness at $z \sim 3.5$  \citep{Richards2006, Worseck2011}, where quasars have colors near that of the stellar locus. The combination of optical and infrared colors allows for more robust classifications, particularly at high-$z$, which is essential for this study (see Section \ref{sec:ttsets}).  

In this paper, we measure the clustering strength of photometrically-selected quasar candidates. We compare these measurements to the theoretical predictions for DM clustering to draw inferences on various physical parameters such as DM halo mass and AGN feedback mechanisms in the early Universe. In Section~\ref{sec:2} we discuss the data used in this study, as well as the techniques to select quasar candidates. Section~\ref{sec:3} provides further details about the two-point autocorrelation function definition and uses. We present our results in Section~\ref{sec:4} and discuss the implications of our results, comparing to several quasar feedback models in Section~\ref{sec:5}. We summarize and conclude in Section~\ref{sec:6}. The Appendices give further relevant and supplemental information. Throughout this paper, we assume a spatially flat $\Lambda$CDM model, consistent with the latest Cosmic Microwave Background \citep[CMB;][]{Planck2016} and Baryon Acoustic Oscillations \citep[BOSS;][]{Alam2016} datasets:  $\Omega_{\rm m} = 0.275$, $H_{0} = 70$ km s$^{-1}$ Mpc$^{-1}$ and $\sigma_{8}=0.77$, unless otherwise stated. All colors and magnitudes in this data set were corrected for Galactic extinction using the parameters (for R$_V$ = 3.1) given in Table 6 of \citet{Schlafly2011}. We calculate magnitudes on the AB scale, which has a flux density zeropoint of 3631 Jy \citep{OkeGunn1983}.

\section{Data and Selection} \label{sec:2}

In this section we describe our datasets including the SpIES and SHELA surveys and the optical data on SDSS Stripe 82. Following that, we describe our `test' and `training' sets required to classify our data. Finally, we present the classification algorithms. The final sample of quasar (candidates) we generate is given in Section~\ref{sec:clustering_sample}.

\subsection{SpIES, SHELA and SDSS Stripe 82}

\begin{figure*}[ht!]
  \centering
  \includegraphics[trim = 0 0 0 225, clip, width = \textwidth]{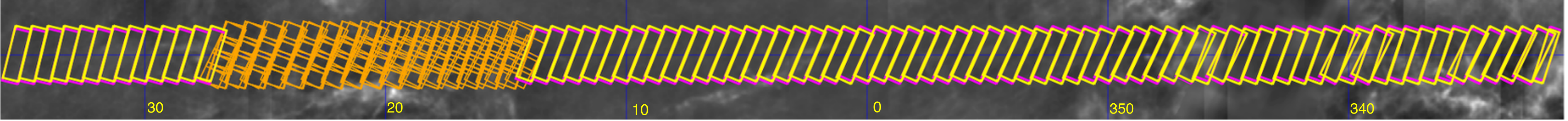}
  \caption{\footnotesize{Superimposed on the 100$\mu$m IRAS dust map \citep{Schlegel1998}, we show the mid-infrared coverage mask on S82 from the SHELA (orange squares) and SpIES (yellow/purple rectangles) survey. These surveys cover $\sim 120$ square degrees on S82 (approximately on third of the full area) and are deep enough to detect quasars out to $z = 6$. Each SpIES observation (individual yellow/purple rectangle) spans a range of 0.82$^{\circ}$ in RA (horizontal axis) and 2$^{\circ}$ in DEC (vertical axis), covering an area of $\sim$ 1.63 deg$^{2}$ each.}}
  \label{fig:S82_footprint}
\end{figure*}

Covering approximately a third of S82 ($-60^{\circ} \leq \alpha \leq60^{\circ}$; $-1.25^{\circ} \leq \delta \leq 1.25^{\circ}$), the SpIES survey was designed to span a large area ($\sim$100 deg$^2$, centered at $\delta = 0$; see Figure \ref{fig:S82_footprint}) and to probe sufficiently deep to select faint, high-$z$ quasars; quasars which were undetected by the Wide-field Infrared Survey Explorer (\emph{WISE}; \citealt{Wright2010}). The SpIES catalogs reported the photometry and photometric errors for $\sim$5.4 million objects at 3.6 $\mu$m and 4.5 $\mu$m. Using SpIES, we are able to detect quasars as faint as $i \sim 22$ with high reliability \citep{Timlin2016}. SpIES is also optimally located to surround existing \emph{Spitzer} data from SHELA \citep{Papovich2016}, forming a long stripe of deep MIR coverage on S82 (see Figure \ref{fig:S82_footprint}).

The SHELA survey was designed to be used alongside the Hobby-Eberly Dark Energy Experiment (HETDEX; \citealt{Hill2008}) to perform dark-energy measurements, requiring deep infrared data. With depths greater than that of SpIES, SHELA provided an additional $\sim$24 deg$^{2}$ of deep infrared coverage on S82 (see Figure \ref{fig:S82_footprint}). In total, SHELA detected $\sim$2 million objects down to \emph{Spitzer} magnitude depths of [3.6]=22.0 and [4.5]=22.6 (compared to 21.9 and 22.0, respectively, for SpIES). In tandem, SpIES and SHELA provide $\sim$120 deg$^{2}$ (accounting for overlapping coverage: Figure \ref{fig:S82_footprint}) of deep, MIR data on S82; data necessary to, along with optical colors, select faint, high-$z$ quasars.

Optical photometric data come from the full SDSS-I/II \citep{York2000} data release as well as the SDSS-III/Baryon Oscillation Spectroscopic Survey (BOSS; \citealt{Eisenstein2011}, \citealt{Dawson2013}). Of particular interest for this study is the S82 coadded catalog (\citealt{Annis2014}; \citealt{Jiang2014}). Imaged with the five optical SDSS filters ($ugriz$; \citealt{Fukugita1996}), S82 was the target for recurring observations to detect variable objects and to obtain deep optical photometry. When the images are stacked, S82 has an optical $i$-band magnitude limit of $i \sim 24.1$ \citep{Jiang2014}, which is significantly deeper than the rest of the SDSS survey. 

Spectroscopically-confirmed quasar data come from the composite quasar catalog of \citet{Richards2015}. This catalog is a compilation of spectroscopic quasars from large surveys such as SDSS (\citealt{York2000}, \citealt{Eisenstein2011}) and the 2QZ project \citep{Croom2004} as well as from smaller surveys such as Hectospec \citep{Fabricant2005}. In total, they compiled $\sim$2 million quasars and quasar candidates (including $\sim$437,000 spectroscopically confirmed quasars) which span a large range in both redshift and $i$-magnitude. The catalog encompasses faint, high-$z$ quasars from BOSS \citep{Paris2014}, which are key to defining the quasar color space used to classify the photometric objects. 

\citet{Richards2015} also matched their catalog to infrared catalogs such as AllWISE\footnote{\url{http://wise2.ipac.caltech.edu/docs/release/allwise/}} and various overlapping \emph{Spitzer} surveys to investigate the mid-infrared colors of these known quasars in the full SDSS field. Mid-infrared color-color diagrams have been particularly useful in quasar classification as shown in \citet{Lacy2004}, \citet{Stern2005}, and \citet{Donley2012}, among others. The addition of this mid-infrared data in classification allows for higher number densities of detected quasars, particularly at high-$z$ ($z \geq$ 2.9).

The new infrared SpIES and SHELA surveys provide a much larger area where deeper infrared data overlaps the optical, providing the necessary information to classify objects as type-1 quasars; the challenge becomes selecting a clean sample of quasar candidates. However, using the machine-learning techniques demonstrated in \citet{Richards2015}, selection of high-$z$ quasar candidates has become much more complete. To generate a final catalog of high-$z$ quasars, we must first assemble a complete sample of all detected objects (i.e.,\ photometric and spectroscopic) to form the \emph{test} set. This test set is then reduced to a subset containing the known (spectroscopically confirmed) high-$z$ quasars used to define the color spaces that train the algorithms along with a fraction of the unknown (photometric) objects (the \emph{training} set). Test objects are then fit using the trained algorithm and are assigned a classification. Presented in Table \ref{tab:datasets} are the demographics of the test and the training sets used in this study, as well as the final selected type-1 quasar candidates.
  
\subsection{Test and Training Sets}\label{sec:ttsets}

In this study, the set of objects to be classified (the test set) and the objects used to train the algorithms (the training set) were constructed in much the same way as in \citet{Richards2015}. The full test set was built using matched optical+MIR photometric data spanning the full SDSS footprint, where \emph{WISE} photometry (converted to \emph{Spitzer} magnitudes) was used when \emph{Spitzer} data did not exist. Furthermore, to be considered for classification, these objects were required to be SDSS sources with $m_{AB}>15$ in all optical bands (to remove contamination due to saturation) and to have the `good' SDSS flags as described in \citet{Richards2015}. The full test set contains $\sim 50$ million objects spanning the full SDSS footprint and includes both spectroscopic and photometric quasars. For this study we further restricted our final test set to objects \emph{only} in S82 since we were particularly interested in candidates where deep \emph{Spitzer} data exists from SpIES and SHELA. After this cut, the final S82 test set is comprised of $\sim2$ million objects with optical+MIR color information. 

\citet{Scranton2002} demonstrated that SDSS star-galaxy separation is relatively clean to $r\sim$21. The Stripe 82 catalogs are \emph{catalog} co-adds, so the deeper data does not yield improved star--galaxy separation without further work. As our targets are typically $r\sim$22, in Section \ref{sec:clustering_sample} and Appendix \ref{append:contamination} we describe the tests intended to exclude low-redshift galaxies acting as interlopers in our sample.

The quasar training set is constructed by first matching the full test set to the high-$z$ quasars in the \citet{Richards2015} composite catalog. In total, there are 22,737 high-$z$ ($z \ge 2.9$) matches between these two sets which we use to train our algorithms, the majority of which come from SDSS (DR7, DR10, and DR12). We also include 12 high-$z$ ($z \geq$ 3.7) spectroscopic quasars from VVDS and McGreer et al. (2013) which were confirmed after the composite catalog was generated. To ensure that the training objects are not confused with other low-$z$ sources, we queried the NASA/IPAC Extragalactic Database\footnote{\url{http://ned.ipac.caltech.edu/forms/nnd.html}} (NED) to check the redshifts. We performed a follow up visual inspection of the spectra for the objects which NED reported to be low-$z$, and removed four objects that had non-quasar spectra. In all, there are 22,745 quasars with $z \geq 2.9$ in the quasar training set to train our machine-learning algorithms.

\begin{figure*}[ht!]
  \centering
  \includegraphics[scale=0.7]{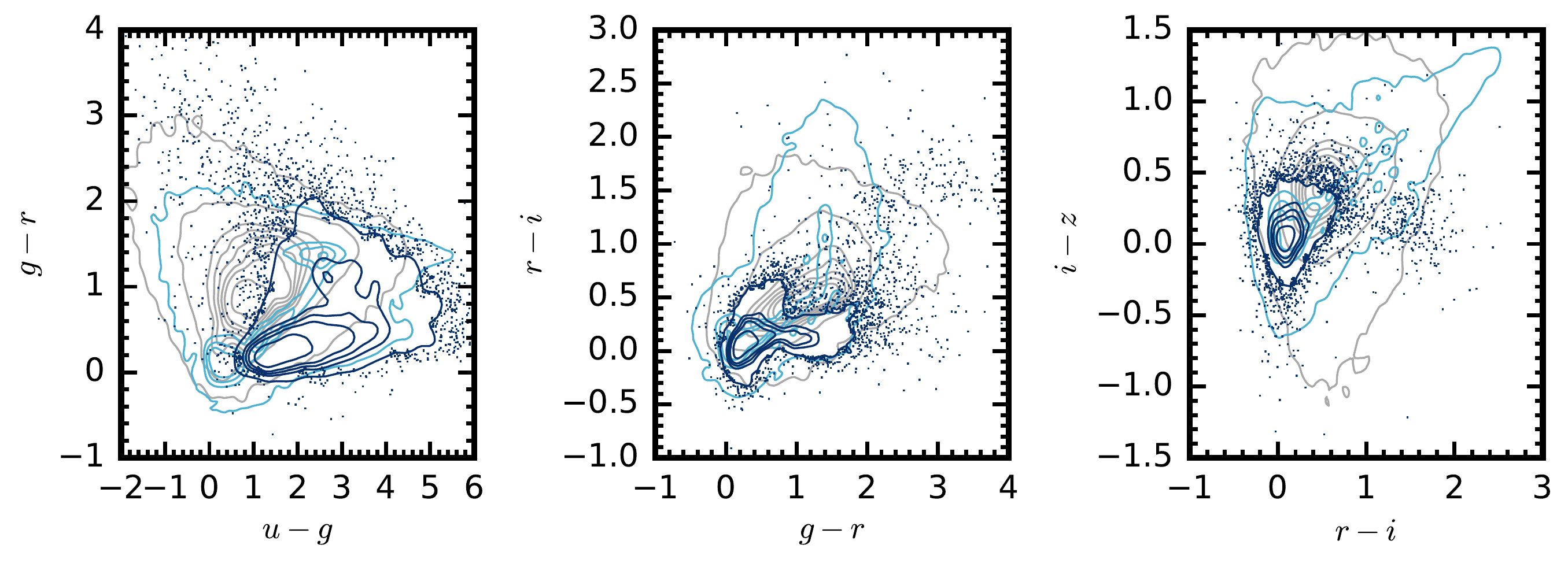}
  \includegraphics[scale=0.7]{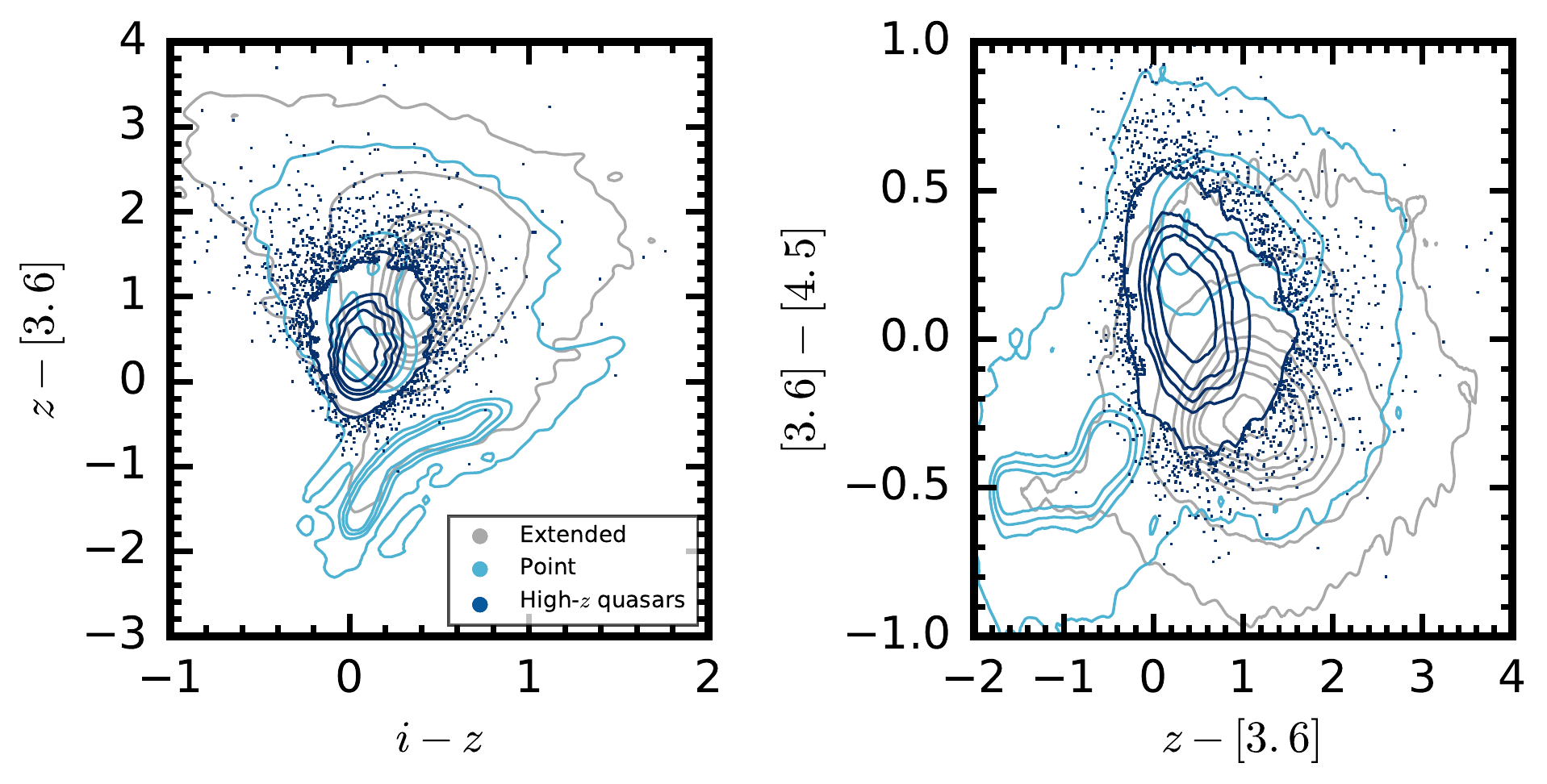}
  \caption{\footnotesize{Optical and infrared colors (computed from AB magnitudes) of the training set objects used to define the classification color spaces. Extended objects in the training set are outlined by the gray contours and point sources are depicted in light blue. The dark blue contours highlight the region where known, high-$z$ (2.9 $\leq z \leq$ 5.2) quasars reside. The overlap of the extended sources and the high-$z$ quasars opens the possibility that we classify, e.g.,\ low-$z$ galaxies as quasars in our algorithms. To remedy this particular situation, we define a metric to identify point sources which eliminates extended object (galaxy) contaminants, and by visual inspection (see Appendix \ref{append:contamination}).}}
  \label{fig:traincol}
\end{figure*}

Additionally, we add to the training set non-quasar sources (`stars'), which do not have spectroscopic information, randomly selected from the full test set. As described in \citet{Richards2015}, the `stars' in the training set can also include previously unclassified quasars, stellar sources, and compact galaxies. The additional `star' information is important in the classification because it defines the color space boundaries around the high-$z$ quasars in the machine-learning algorithms. The full training set is comprised of $\sim${700,000} `stars' and confirmed quasars in the SDSS footprint that are as faint as $i \sim 23$ and observed to $z \sim6$. In this investigation, we split the training set into two redshift ranges; a lower-$z$ ($2.9 \leq z < 3.5$) and a higher-$z$ ($3.5 \leq z \leq 5.2$) range for selection. The colors of the higher-$z$ objects are much more distinct from low-$z$ `stars' compared to the objects in the lower-$z$ range, thus the selection is much more efficient in the higher-$z$ range. Figure \ref{fig:traincol} depicts the colors of extended and point sources in the training set, and highlights the colors of the known high-$z$ quasars in each color space used to classify the test objects. We also provide the demographics for both the testing and training sets in Table \ref{tab:datasets}.

\begin{deluxetable}{lccc}
\tablecolumns{4}
\tablewidth{0pt}
\tablecaption{Training and Test Sets}
\tablehead{
  \colhead{Data Set} &  \colhead{N$_{Obj}$} &  \colhead{$z$-range} &  \colhead{$i$mag}  \\
}
\startdata
 Full Training & $\sim7.00\times10^5$ & $0\leq z\leq6$ & $15\leq i \leq24$  \\
 High-$z$ quasars  & $\sim2.27\times10^4$ & $2.9\leq z\leq6$ & $16\leq i \leq23$ \\
 Full Test & $\sim5\times10^7$  & -- & $16\leq i \leq23$  \\
 S82 Test & $\sim2\times10^6$  & -- & $16\leq i \leq24$ \\
 Candidates$^{a}$  & 1378 & $2.9\leq z^{b}\leq5.1$ & $18\leq i \leq23$  
\enddata
\tablecomments{The training and testing set demographics. The training set is a combination of spectroscopic objects and photometric objects in the full test set. The training quasars is the compilation of spectroscopic quasars used to train the algorithms that are used to classify the S82 test objects. Combining the photometric and spectroscopic quasars, there are 1378 high-$z$ quasars with which to compute the correlation function.\\
$^{a}$ Contains both spectroscopic and photometric quasars\\
$^{b}$Photometric redshifts} 
\label{tab:datasets}
\end{deluxetable}

 \subsection{Classification Algorithms}
The colors of confirmed high-$z$ quasars in the training set (shown in Figure \ref{fig:traincol}) are used to teach the machine-learning algorithms where high-$z$ quasars lie in multi-dimensional color space. Colors of the photometric objects in the S82 test set are then input into the trained algorithms to classify them as high-$z$ quasars. For this analysis, we utilize three classification algorithms; {\it{Random-Forest Classification}} (RF), {\it{Support-Vector Classification}} (SVC), and {\it{Bootstrap Aggregation (Bagging) on K-Nearest Neighbors}} (KNN), which we define below. All of these algorithms are openly available in the Scikit-Learn\footnote{\href{sklearn}{http://scikit-learn.org/stable/}} Python package used in this study. 

The RF classifier\footnote{\href{RFclass}{http://scikit-learn.org/stable/modules/ensemble.html\#forest}} creates a set of $N$ random decision trees, which split the training quasars by their colors into different branches, with each branch returning a classification (in this case high-$z$ or not). The colors of the test objects are then subject to the splitting that each tree has created, and each of the trees assign a classification based on the conditions that the test objects satisfy. The mode result of all of the trees is used as the final classification for each of the test objects.

We also employ the SVC algorithm\footnote{\href{SVC}{http://scikit-learn.org/stable/modules/svm.html\#svm}}, which defines an optimal hyperplane that separates two populations of objects by the largest margin. In this case, the training set objects create the six-dimensional color space, and the hyperplane is defined by the plane that maximally separates the known high-$z$ quasars from the `stars' in the training set. Classification of the test objects is based on the side of the hyperplane they lie in this multi-dimensional color space. 

Finally, we use ``Bagging" with a KNN algorithm\footnote{\href{BKNN}{http://scikit-learn.org/stable/modules/neighbors.html\#classification}}, where Bagging is the process of splitting the training set into $N$ different subsets of randomly chosen training objects (with replacement). Each of those subsets is used to train the machine-learning algorithm (KNN in this case), resulting in $N$ trained KNN algorithms. The KNN algorithm assembles the training set color information and classifies the test data by analyzing the closest `k' training objects in color space. Similar to a majority rule, the test object is classified based on the type of the closest `k' training object (in this case, high-$z$ quasar or not). This analysis is done in all of the Bagging subsets, and the mean result from all of the bags is chosen as the final classification. 
  
To measure the effectiveness of each algorithm, we compute the two key selection parameters: efficiency and completeness. The efficiency of an algorithm relates the number of objects that it classifies correctly to the total number of objects it classifies, and can be used to estimate the contamination of the classified sample by taking the difference from unity. Completeness is a measure of how many quasars are properly classified compared to the total number of known quasars in the data set. 

Estimation of the completeness and efficiency of our algorithms requires the full training set to be split into two subsets for cross-validation (CV); a subset with 75\% of the data to be used as a CV `training set', and a subset with 25\% of the data to be used as CV `test' objects. These sets are input into the classification algorithms discussed above. Since the CV test objects contain known quasars, completeness and efficiency can be calculated using the classification results of the known quasars from the CV test set. Ideally, both completeness and efficiency should be maximized to recover all of the high-$z$ quasars, and only the high-$z$ quasars. Practically, however, quasar colors can overlap with stars and low-$z$ galaxies, so contamination and missed classifications are inevitable. 

We compare our algorithms to the kernel density estimation (KDE) used in \citet{Richards2015}. This study classified photometric objects in the SDSS footprint using optical data along with infrared data from \emph{WISE}. The KDE method used in \citet{Richards2015} first defined a color `bandwidth' for each class of object (quasar or non-quasar) which acts to smooth the color distributions, and a Bayesian stellar prior which defines the percentage of objects in the test sets thought to be `stars' (i.e.,\ non-quasars). A probability density function (PDF) is then defined in color space for a class of object, and the likelihood that a test object with certain photometric colors belongs to a class is computed using the bandwidth and a kernel function. The posterior probability that an object is a quasar given its color is computed by applying Bayes' theorem using the defined priors and likelihoods for each test object (see \citealt{Richards2009} for details). This study classified objects over a wide range of redshifts, however we will compare the performance of our algorithms to their highest redshift classification ($3.5 \leq z \leq 5$). 

In our investigation, we split the classification of quasar candidates into a lower-$z$ (2.9$\leq z <$ 3.5) and a higher-$z$ ($z \geq$ 3.5) bin. We found that the classification algorithms performed better at higher-$z$ compared to lower-$z$ as reported in Table \ref{tab:classifications}. This mainly because quasars begin to drop out of the SDSS $u$-band filter at $z \sim$ 3.5, which significantly alters the $u-g$ color space and helps the machine-learning algorithms efficiently select these objects. The colors of $z \sim$ 3 quasars are very similar to those at $z \sim$ 2.2, therefore the algorithms tend to confuse low-$z$ quasars with higher redshift quasars. Through cross validation, we found that the ``Bagging" algorithm performed the best at lower-$z$ and all three perform equally well at higher-$z$ as reported in Table \ref{tab:classifications}. More details are presented in Section \ref{sec:clustering_sample} where we describe our final candidate selection.


 \begin{deluxetable}{lccc}
\tablecolumns{4}
\tablewidth{0pt}
\tablecaption{Estimated Completeness and Efficiency}
\tablehead{
  \colhead{Algorithm} &  \colhead{Completeness} &  \colhead{Efficiency} &  \colhead{Contamination}  \\
  \colhead{} &  \colhead{\%} &  \colhead{\%} &  \colhead{\%}
}
\startdata
 RF & $83$/$78$/$80$ & $43$/$93$/$86$ & $57$/$7$/$14$ \\
 SVC & $82$/$79$/$79$ & $40$/$95$/$86$ & $60$/$5$/$12$ \\
 Bagging KNN & $83$/$80$/$80$ & $85$/$95$/$88$ & $15$/$5$/$12$ \\
 KDE & $-$/$78/-$ & $-$/$97/-$ & $-$/$3$/$-$ 
\enddata
\tablecomments{Estimated completeness, efficiency and contamination measured for the three algorithms used in this study compared to the KDE method used in \citet{Richards2015}. The first three rows report our algorithms when selecting in a lower redshift range ($2.9 \le z <3.5$; left), a higher redshift range ($3.5 \le z \le 5.2$; center), and when selecting in a broader redshift range ($2.9 \le z \le 5.2$; right). The values in the center are used to compare to \citet{Richards2015}. These values are estimates since the actual test set probes slightly fainter than the validation set.}
\label{tab:classifications}
\end{deluxetable}

\subsection{Photometric Redshifts}\label{sec:photz}

Photometric redshifts of our candidates were estimated with Nadaraya-Watson (NW) kernel regression\footnote{\href{NWreg}{http://www.astroml.org/modules/generated/\\astroML.linear\_model.NadarayaWatson.html}}. NW is a natural extension of more familiar regression techniques. Linear regression fits a line to 2-D data. Polynomial regression instead fits a higher order curve. Basis function regression (of which polynomial regression is an example) uses a pre-determined ``basis'' function to fit the data. NW is just basis function regression using a Gaussian kernel \citep{Ivezic2014}.

The NW algorithm defines the multi-dimensional color space of the training objects with spectroscopic redshifts, then builds a kernel matrix, $K$, which measures the pairwise distance between the colors of the test objects and the colors of the training objects, where $K$ is the Gaussian kernel:
\begin{equation}
K = \exp \left( \frac{1}{2\sigma^2} \Vert d_{\rm test}-d_{\rm train}\Vert ^2\right). 
\label{eqn:gsnkrnl}
\end{equation}
Here, $\Vert d_{\rm test}-d_{\rm train}\Vert$ is the Euclidean distance between the colors of the test objects and the color of the training objects, and $\sigma$ is the bandwidth of the kernel ($\sigma= 0.05$ produced the best self-validation results in this study). From Equation \ref{eqn:gsnkrnl}, if a test object is close to a training object (i.e.,\ if the 6D colors are very similar), the kernel approaches 1; however, the further the colors are from each other, the smaller the Gaussian kernel becomes. Therefore, the kernel matrix is used as weights in the estimate of the photometric redshift, defined by:
\begin{equation}
z_{\rm phot} = \frac{\sum_i K_i \cdot z_{{\rm spec}, i}}{\sum_i K_i},
\end{equation}
where the kernel element in $K$ is multiplied by the spectroscopic redshift corresponding to the training quasar input into Equation \ref{eqn:gsnkrnl}. The final photometric redshift result for a candidate object is then the weighted sum over all the spectroscopic redshifts of the training objects.

\begin{figure}[h!]
 \centering
 \includegraphics[scale=0.33]{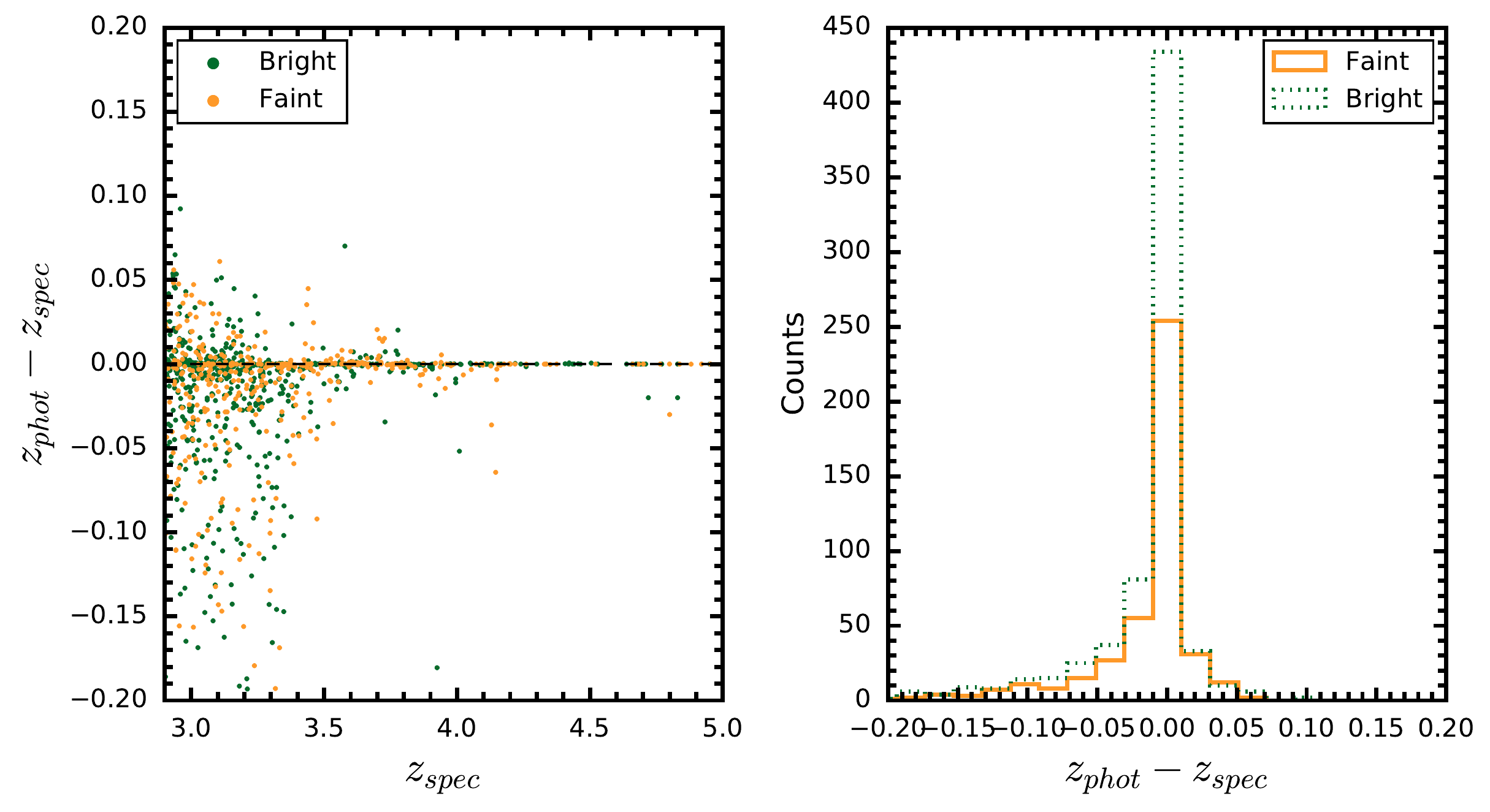}
 \caption{\footnotesize{Left: Comparison of known high-$z$ quasar spectroscopic redshifts with the estimated photometric redshift using Nadaraya-Watson regression. The known quasars are split into bright ($i\leq$ 20.2; green points) and faint ($i>$ 20.2; orange points) bins to test the effectiveness of this algorithm for quasars of different brightness. The black dashed line depicts $z_{phot} - z_{spec} = 0$. Right: Difference between the photometric and spectroscopic redshifts for the bright and faint quasars. Approximately $\sim93\%$ of the high-$z$ quasars are constrained to $\vert \delta z \vert \leq 0.1$ in both bins.}}
 \label{fig:photvspec}
 \end{figure}
 
To test the effectiveness of this method, we calculate the photometric redshifts of the spectroscopic quasars on S82 over all redshift ranges using the same training set we use for the candidates. Additionally, we split the quasars into a bright and faint subset, where we differentiate between bright and faint at $i$=20.2. The results in Figure \ref{fig:photvspec} show that there is a tight correlation between the spectroscopic redshift of the quasar and its estimated photometric redshift for both subsets. In both cases $\sim 93\%$ of the photometric redshifts differ from the spectroscopic redshifts by no more than, $\vert \delta z \vert \leq 0.1$. These results are similar to the findings in \citet{Richards2015} for their highest redshift bin, who used an empirical method outlined in \citet{Richards2001} and \citet{Weinstein2004}. 

\begin{figure}[h!]
 \centering
 \includegraphics[scale=0.45]{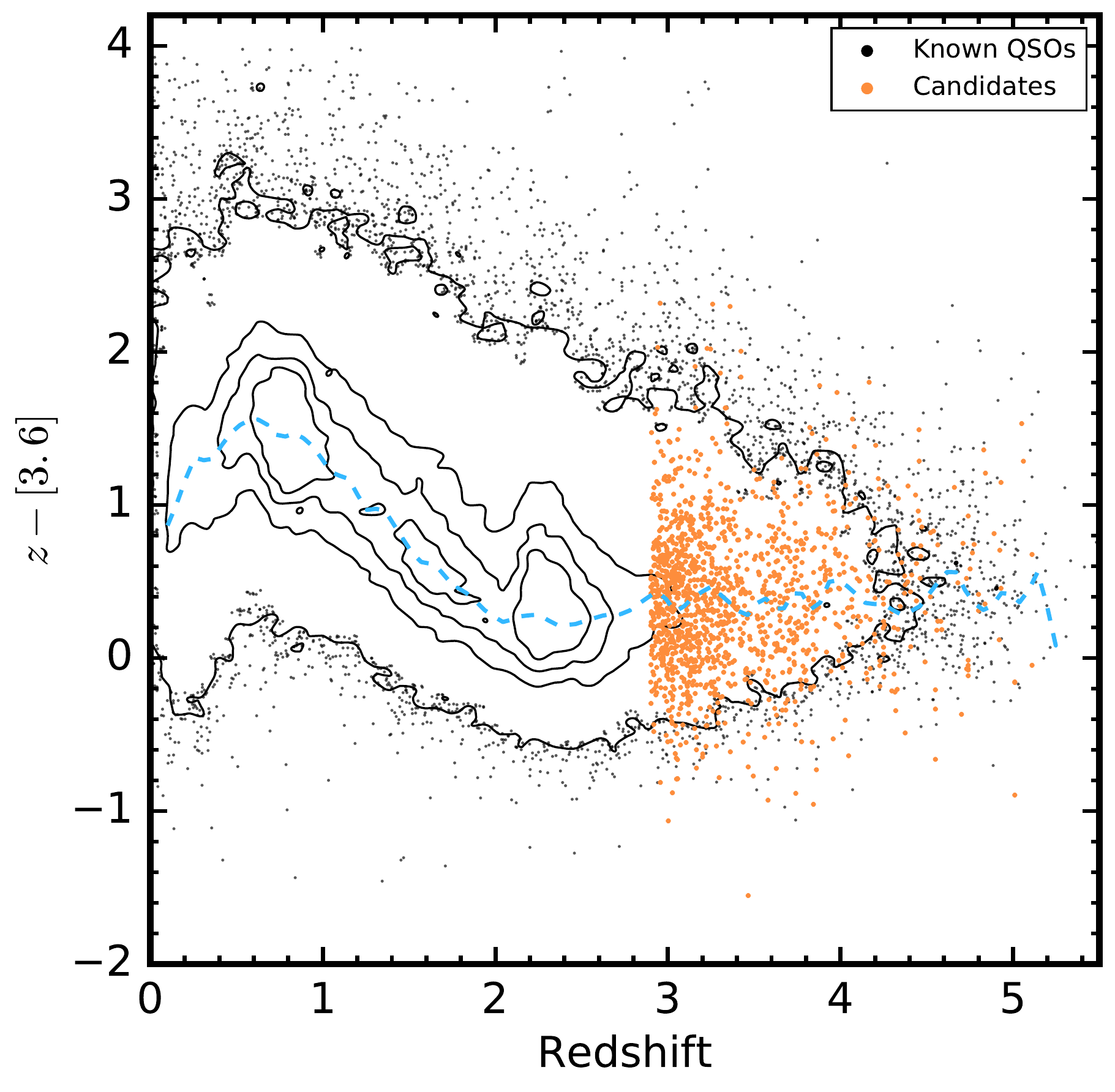}
 \caption{\footnotesize{Example color-redshift diagram of the spectroscopic training data (black) and the photometric redshifts of the candidates (orange). The photometric redshifts estimated using the NW algorithm share the same color space as the spectroscopic sample on which the algorithm was trained ($2.9 \leq z \leq 5.1$). The blue dashed curve indicates the modal color as a function of redshift of the known quasars.}}
 \label{fig:colzspecsing}
 \end{figure}

Using the NW regression algorithm, each candidate quasar selected with the aforementioned algorithms was assigned a photometric redshift. A comparison of the candidate redshifts to the spectroscopic redshifts is displayed in Figure \ref{fig:colzspecsing}. With candidates selected, and their photometric redshifts computed, we now create a final sample of candidates with which to compute the correlation function.

 \subsection{Clustering Sample}\label{sec:clustering_sample}
 
 \begin{figure*}[ht!]
  \centering
  \includegraphics[scale=0.7]{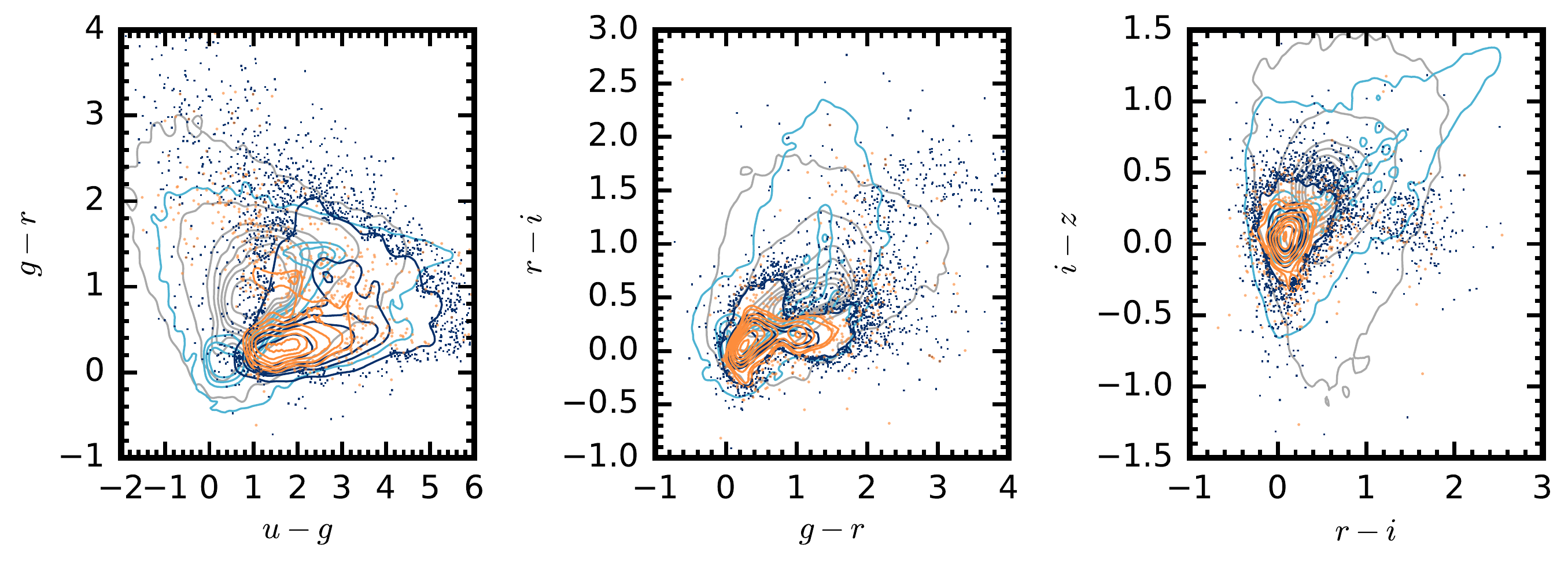}
  \includegraphics[scale=0.7]{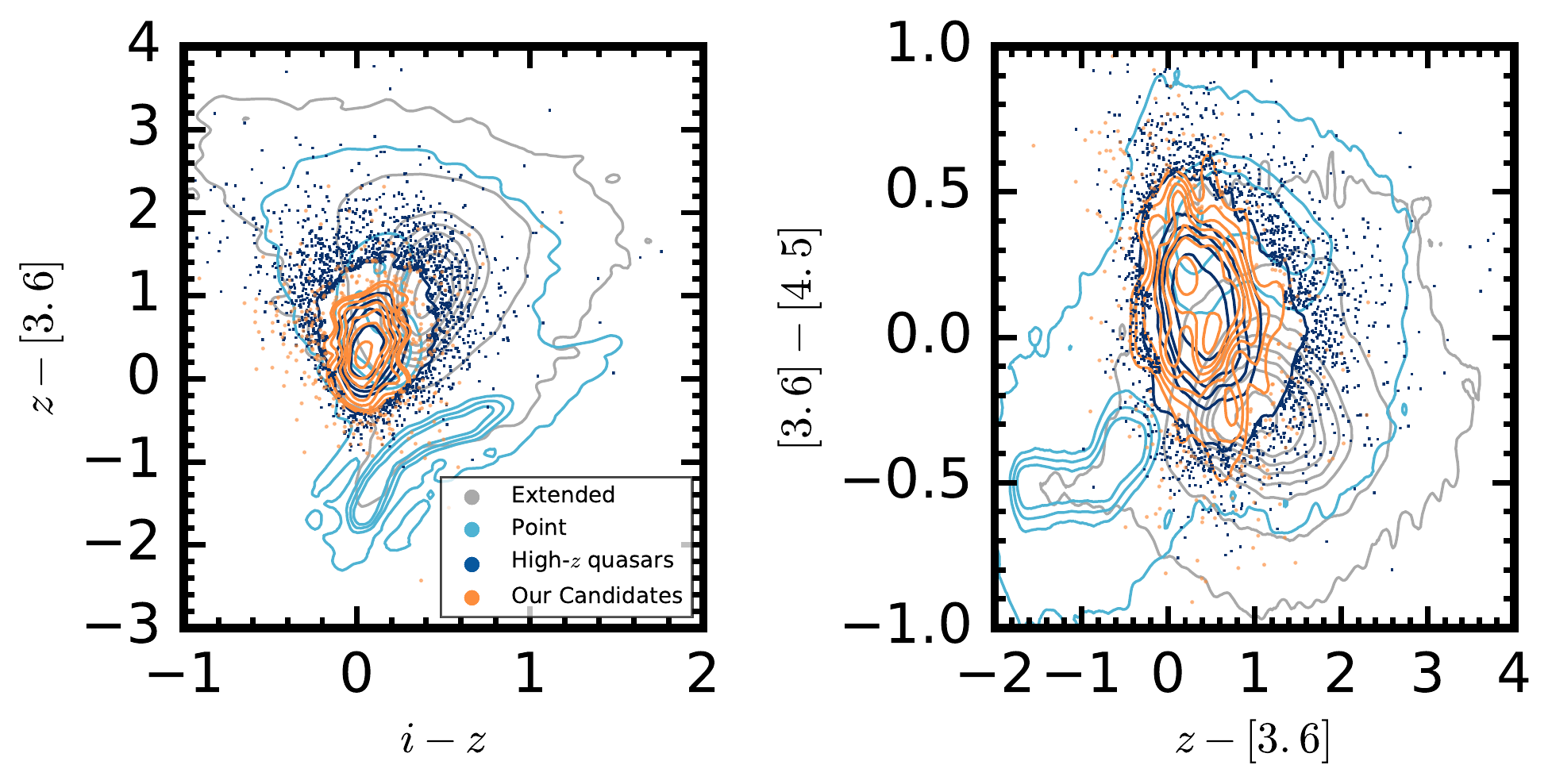}
  \caption{\footnotesize{Optical and infrared colors of the selected quasars (orange contours). The other contour colors are the same as in Figure \ref{fig:traincol}. These panels demonstrate that the location of the candidates in color space overlap with the colors on which they were trained (dark blue contours).}}
  \label{fig:candcol}
\end{figure*}
 
Although classification was performed on all of the S82 test objects, and photometric redshifts were computed for all candidates that were selected, we further restricted the data set to create the cleanest sample of faint, high-$z$ quasars with which to compute the 2PCF. First, to retain the faintest objects with the deepest photometry, we required that the objects lay within the SpIES/SHELA footprint, where the deep MIR data exists, \emph{and} that they were sufficiently far away from bright stellar sources which contaminate the photometry (see \citealt{Timlin2016} for more details). Additionally, candidates were required to have photometric redshifts in the range $2.9 \leq z \leq 5.1$, enabling us to compare our results with \citet{Shen2007}: the most recent wide-area spectroscopic study of quasar clustering at redshifts as high as $z \sim 4$. 

To ameliorate potential sources of contamination, yet to select as many true high-$z$ quasars as possible, we combined the results of each of the selection algorithms (see Table \ref{tab:classifications}). At low-$z$ ($2.9 \leq z < 3.4$), we chose to \emph{only} employ the ``Bagging'' classifier because of its high efficiency. While including the results from the other two classifiers would have made our sample more complete, it also would have added a large amount of contamination. At high-$z$ ($3.4 \leq z \leq 5.2$), however, we combined the selection results of the three algorithms since they all have low contamination as shown in Table \ref{tab:classifications}. 

Despite combining the classification results in this manner, the sample still contained contamination from low-$z$ galaxies. To eliminate the obvious galaxies, we restricted our data to point-like sources only. We generated our own metric for high-$z$ quasar point sources by taking the difference between the {\tt PSFMAG}\footnote{\label{notemags}\url{http://www.sdss3.org/dr10/algorithms/magnitudes.php}} and {\tt cMODELMAG}\textsuperscript{\ref{notemags}} ($\delta_{mag}$) in the SDSS DR10 $i$-band. A difference of $\delta_{mag}\leq 0.145$ is used in the SDSS catalogs to label an object as a point source. We found that the known quasars in our lower-$z$ range ($2.9 \leq z < 3.4$) had $\delta_{mag}=0.2$, whereas the known higher-$z$ ($z \geq 3.4$) had $\delta_{mag}=0.15$. We apply this morphology cut to the selected objects in the appropriate redshift range after the selection had been performed. This cut eliminated a significant fraction of extended sources, which we consider to be contaminants in our sample (confirmed using visual inspection; see Appendix \ref{append:contamination}).

Another source of contamination that we account for is high Galactic extinction objects which can cause low-$z$ objects to be mistaken for high-$z$ quasars \citep{Myers2006}. Removal of these highly extincted objects is particularly important in this study since the eastern edge of the SpIES field overlaps with the Galactic plane ($330^{\circ} \leq \alpha_{J2000} \leq 344.4^{\circ}$). To remove the contamination due to these objects, we elect to cut out this region from our final analysis (see Appendix \ref{append:contamination} for more details). While this process eliminates some area over which we can perform the clustering analysis, it also removes contaminants that are confused for high-$z$ quasars in the machine-learning algorithms (despite the extinction-corrected magnitudes).

Differences in the angular mask of the data and the randoms can also affect our clustering measurement. The edges of the SHELA field are not uniformly covered in the mask, requiring that we cut in declination ($-1.2^{\circ} \leq \delta_{J2000} \leq 1.2^{\circ}$) to ensure that the densities of the data and randoms were approximately the same across the field. After cutting out the extinction region and these under-dense regions, our final footprint covers 102 deg$^2$ on Stripe 82.

Finally, every candidate object (before and after the morphology cut) was visually inspected using the stacked $g$, $r$, $z$ images from the Dark Energy Camera Legacy Survey (DECaLS\footnote{\url{legacysurvey.org}}) image cutout tool\footnote{\url{https://github.com/yymao/decals-image-list-tool}}. DECaLS images to similar depths as the SDSS S82 coadded catalog ($r=$ 23.4 compared to $r=$ 24.6 on S82), but uses the Dark Energy Camera (DECam\footnote{\url{http://www.ctio.noao.edu/noao/content/DECam-User-Guide}}), which has a finer resolution than SDSS (0.26$\arcsec$ compared to 0.39$\arcsec$ per pixel). This added resolution enabled us to visually eliminate a small number of obvious low-redshift galaxies which share color spaces with high-$z$ quasars (see examples in Appendix \ref{append:contamination}). 

After the cuts and visual inspection, 1378 objects remained as high-$z$ quasars (see Table \ref{tab:candidates}). Of these, 726 are spectroscopically confirmed from the \citet{Richards2015} comprehensive catalog and we select 652 new high-$z$ quasar candidates with which we can measure the 2PCF. None of the quasars or candidates used in this study were used in the \citet{Shen2007} study. The colors of our selected quasars are presented in Figure \ref{fig:candcol} and compared to the colors of the training objects. The majority of the selected quasars share the same color space as the high-$z$ quasars whose colors were used to train the algorithms, but as our candidates delve fainter than the majority of the training objects, there is some scatter in their colors. While there is stellar contamination in the sample (which we will model in Section \ref{sec:4}), some of the scatter in the colors could be due to contamination from objects such as compact galaxies, which are more difficult to identify from colors alone. 

\begin{figure}[h!]
 \centering
 \includegraphics[scale=0.5]{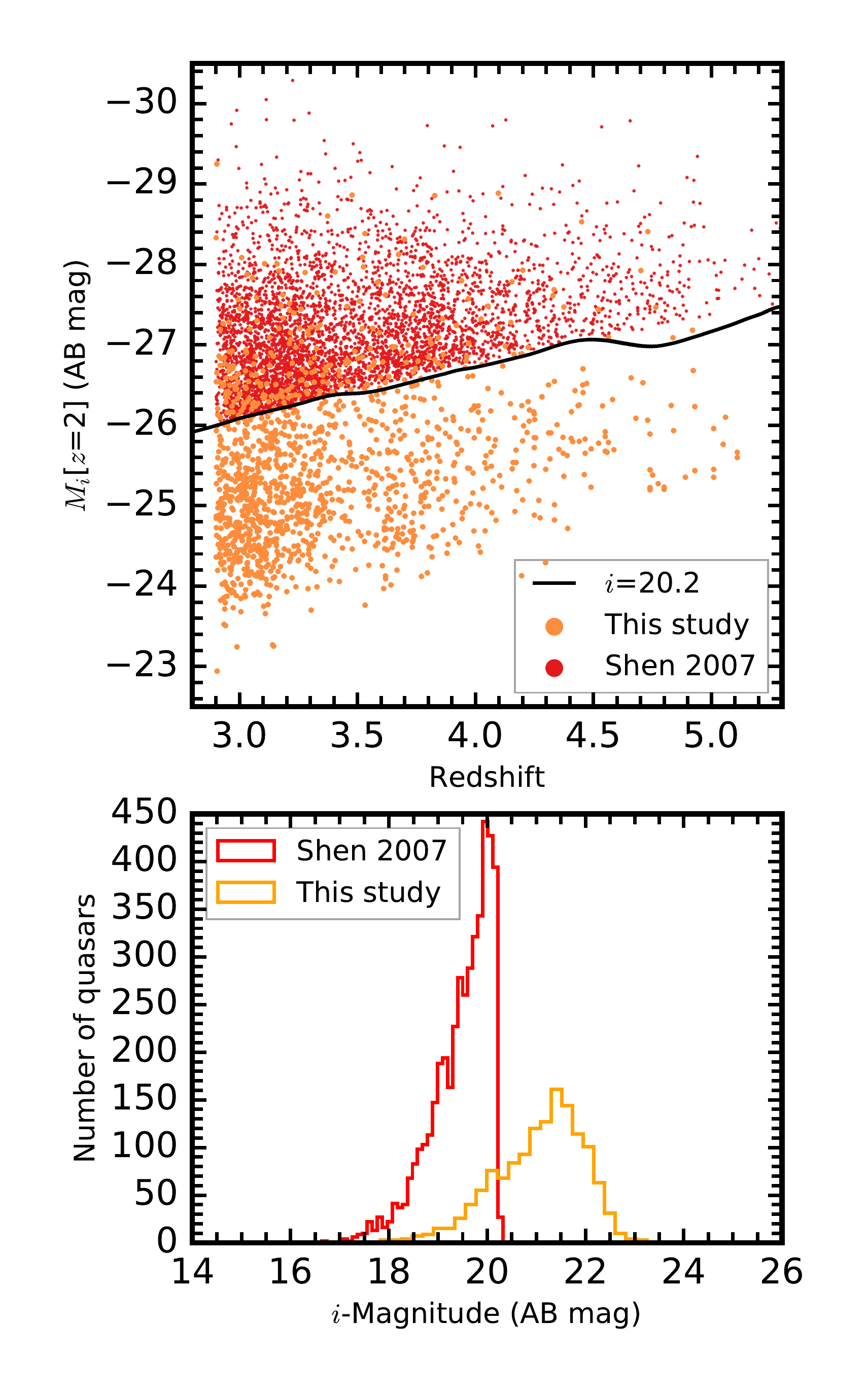}
 \caption{\footnotesize{Top: Absolute $i$-magnitude of the quasar candidates (orange points) compared to the spectroscopically-confirmed quasars from \citet{Shen2007} (red points). The solid black curve depicts constant $i$-magnitude ($i=20.2$), where the \citet{Shen2007} objects are brighter than this magnitude and the photometric candidates are fainter. The $i$-magnitudes were corrected for reddening to $z=2$ using the model from \citet{Richards2006}. Bottom: Distribution of $i$-magnitudes for our candidates (orange) compared to the \citet{Shen2007} candidates (red). We have far fewer bright objects ($i\leq$ 20.2) because our survey area is much smaller.}}
 \label{fig:Abs_mag}
 \end{figure}

Using the redshifts and the $i$-band apparent magnitudes, we compute the absolute magnitude of these quasar candidates, and compare them to the spectroscopic sample from \citet{Shen2007} (renormalized to $z=2$ after K-correcting using the model in \citealt{Richards2006}) in the top panel of Figure \ref{fig:Abs_mag}. The majority of the photometric candidates are fainter than the \citet{Shen2007} quasars and are fainter than $i=20.2$ (shown in the bottom panel of Figure \ref{fig:Abs_mag}), {\it{which is necessary to break the degeneracy in the bias as a function of redshift}}. This investigation contains a small number of objects brighter than $i=20.2$ compared to \citet{Shen2007} because it covers a smaller area ($\sim$ 100 deg$^2$ vs.\ $\sim$ 4000 deg$^2$, respectively).
 
\begin{deluxetable*}{ll ccccccccccc}[h!]
\tablecolumns{13}
\tablewidth{0pt}
\tablecaption{Quasar Candidate Table}
\tablehead{
  \colhead{$\alpha_{J2000}$} &  \colhead{$\delta_{J2000}$} &  \colhead{Au} &  \colhead{$ug$} &  \colhead{$gr$} &  \colhead{$ri$} &  \colhead{$iz$} &  \colhead{$zs1$} & \colhead{$s1s2$} & \colhead{$i$mag}  & \colhead{$z$spec} & \colhead{$z$photNW} & \colhead{$z$best}  \\
  \colhead{(degrees)} &  \colhead{(degrees)} &  \colhead{} &  \colhead{} &  \colhead{} &  \colhead{} &  \colhead{} &  \colhead{} &  \colhead{} &  \colhead{(AB; asinh)} &  \colhead{} &  \colhead{} &  \colhead{}
}
\startdata
 31.92786 &	-0.04275 &	0.15 &	2.17 &	0.933 &	0.277 &	0.203 &	0.811 &	0.329 & 21.00 &	-999.0 & 3.89 &	3.89 \\
12.30498 &	0.69156 &	0.14 &	0.23 &	2.04 &	0.113 &	0.030 &	0.631 &	-0.543 &	22.10 &	-999.0 &3.90 & 3.90 \\
27.56491 &	0.76547 &	0.14 &	1.84 &	1.50 &	0.205 &	0.283 &	0.600 &	0.143 &	20.76 &	3.90 &3.91 &	3.90 
\enddata
\tablecomments{List of all candidates selected by the three algorithms. Along with positional information, we record the $i$-band AB magnitude, its $u$-band extinction parameters, and the optical/infrared color of each object. We also report the spectroscopic redshift if the quasar is a confirmed object (a value -999.0 indicates that there is no spectroscopic redshift). The photometric redshift estimate from the Nadarya-Watson regression algorithm is recorded in the next column followed by the `best' redshift estimate (records the spectroscopic redshift instead of the photometric estimate, when available). The full version of this catalog can be found at \url{https://github.com/JDTimlin/QSO_Clustering/tree/master/highz_clustering/clustering/Data_Sets}.}
\label{tab:candidates}
\end{deluxetable*}


\section{Clustering}\label{sec:3}
 \subsection{Two-Point Correlation Function}
 Spatial clustering of a population of objects is quantified using the 2PCF, which is the joint probability of finding an object in two volume elements, $dV_1$ and $dV_2$, at some separation $\mathbf{r_{12}}$ \citep{Peebles1980}. This quantity can be expressed as:
 \begin{equation}\label{1980eqn}
 dP = n^2[1+\xi(\mathbf{r_{12}})]\ dV_1\ dV_2
 \end{equation}
where $n$ is the mean number density and $\xi(\mathbf{r_{12}})$ is the correlation function. In this equation, if the 2PCF is zero, the probability shows no excess compared to a Gaussian random distribution. We can derive this statistic for a distribution of objects in a density field, $\rho$, where the probability of finding an object in that field is $dP = \langle\rho(r)\rangle\ dV$ \citep{Peebles1980}. The probability of finding a pair of objects in two density fields $\rho_1, \rho_2$ separated by a distance $r$ is:
 \begin{equation}\label{denprob}
 dP = \langle\rho_1(r)\rangle \langle\rho_2(r)\rangle\ dV_1\ dV_2
 \end{equation}

The density in an expanding Universe is modeled with a linear perturbation $\rho(r) = \bar{\rho}[1+\delta(r)]$, so Equation \eqref{denprob} becomes:
 \begin{equation}
  dP = \langle\bar{\rho}[1+\delta_1(r')]\rangle \langle\bar{\rho}[1+\delta_2(r)]\rangle\ dV_1\ dV_2 
 \end{equation}
 \begin{equation}
  dP = \bar{\rho}^2 [1+\langle \delta_1(r') \delta_2(r)\rangle]\ dV_1\ dV_2
 \end{equation}
 Comparing with Equation \eqref{1980eqn} we see that the correlation function is the ensemble average of the perturbations, $\xi(r_{12}) = \langle \delta_1(r') \delta_2(r)\rangle$. The density field can also be expressed in Fourier space \citep{Bonometto2002}:
 \begin{equation}
 \delta(r) = \frac{1}{(2\pi)^3}\int \delta(k) e^{-ikr}dk.
 \end{equation}
Taking the Fourier transform of the correlation function yields:
 \begin{equation}
 \langle \delta_1(r') \delta_2(r)\rangle = \frac{1}{(2\pi)^3}\int \langle\delta(k)\delta^*(k)\rangle e^{-ikr}dk
 \end{equation}
where the ensemble average of the density modes, $\langle\delta(k)\delta^*(k)\rangle$, is the definition of the power spectrum, $P(k)$. The correlation function is, therefore, the Fourier transform of the power spectrum.  We relate our clustering results to the theoretical clustering of DM, which will be obtained through calculation of the DM power spectrum. In this paper, we compute the angular projected correlation function, $\omega(\theta)$, which is a projection from three dimensional (3-D) volume space into two dimensional (2-D) angular space.
 
 \subsection{Estimating the Correlation Function}
To estimate the correlation function, one needs to compare the data set to a set of randomly distributed points. To compute the correlation function we use the estimator from \citet{Landy1993}:
 
 \begin{equation}\label{LSest}
 \omega(\theta) = \frac{\langle DD \rangle - 2 \langle DR \rangle + \langle RR \rangle}{\langle RR \rangle}
 \end{equation}
where $\langle DD \rangle, \langle DR \rangle$, and $\langle RR \rangle$ are the data-data, data-random, random-random pair counts within an angular separation of $\theta$ (to measure the three dimensional correlation function, $\xi(s)$ one simply counts pairs within a 3-D comoving separation distances). The pair counts are normalized by the ratio of the number of objects in the data and random sets. To reduce the shot noise in the measurement, we use $\sim$100 times the number of points as the data catalog. The normalization of the pair counts reconciles that there are more random points to match than data. 

\begin{figure}[h!]
\centering
\includegraphics[scale = 0.45]{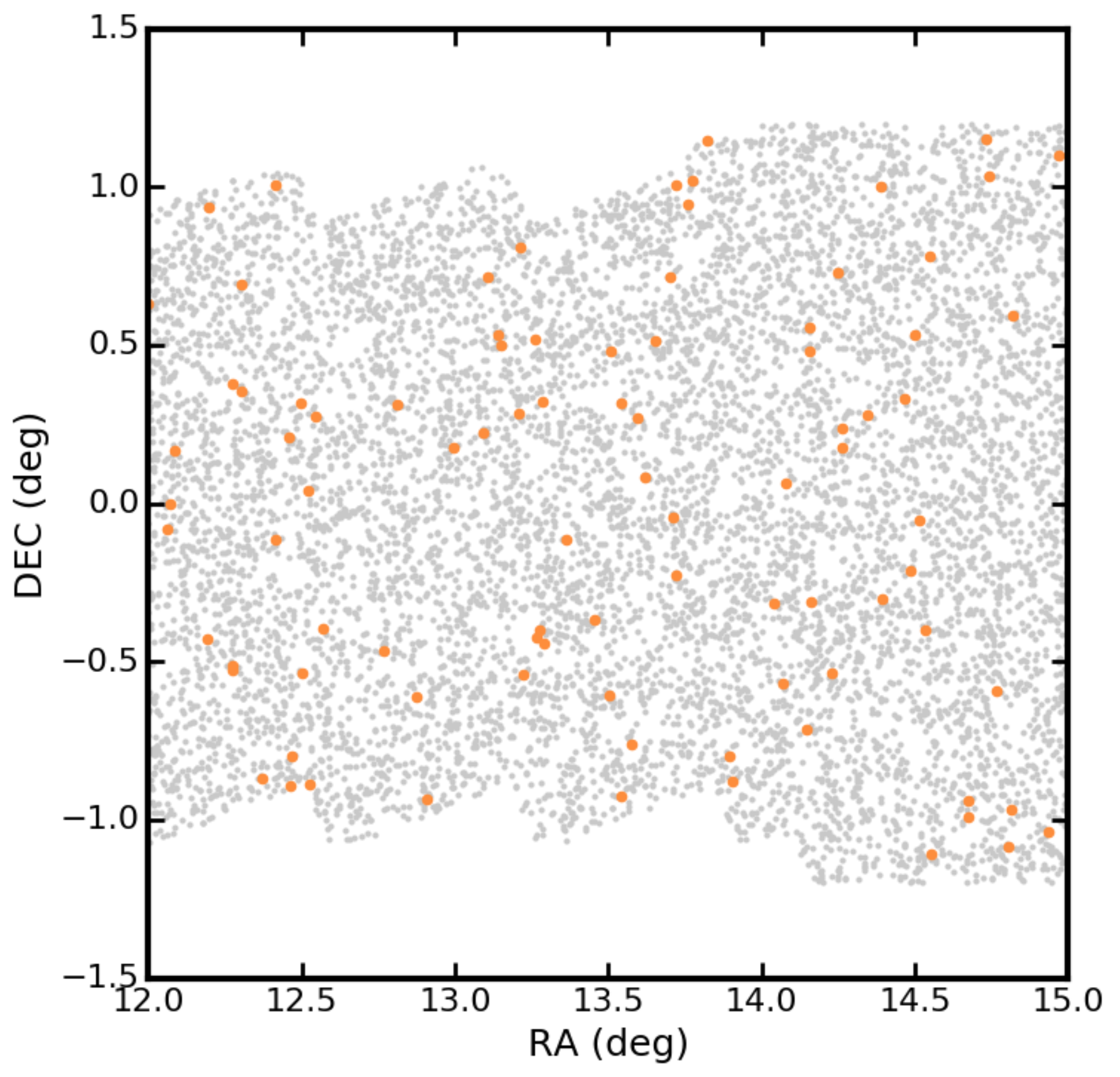}
\caption{\footnotesize{High-$z$ quasar data set (orange) and the random mask (grey) used to perform the clustering analysis. The holes in the mask are cutouts of bright stars in the SpIES and SHELA field where the radius of the hole corresponds to the brightness of the star (see \citealt{Timlin2016} for more details). The holes and corners in this mask identify locations where candidates cannot be selected; to eliminate bias, we mask these regions using MANGLE. Additionally, we exclude objects in the declination range $-1.2^{\circ} \leq \delta_{J2000} \leq 1.2^{\circ}$ due to coverage issues in the edges of the SHELA fields.}}
\label{fig:rndmsk}
\end{figure}

The random data must lie on an identical angular mask as the data. To generate the random catalog for our candidates, we first construct the angular mask using the MANGLE\footnote{\url{http://space.mit.edu/~molly/mangle/}} software \citep{Swanson2008}. This package allows a user to combine polygons from telescope observations to create a continuous mask, with accurate boundaries and holes, on the surface of a sphere. We combine the fields from the SpIES and SHELA surveys (see Figure \ref{fig:S82_footprint}), and remove circular regions of varying radii around bright stars from the 2MASS Point Source Catalog as outlined in \citet{Timlin2016}. Objects in these regions were excluded from the selection of quasars because they are contaminated by the excess flux from the bright star, so we mask them using MANGLE. Random positions are chosen across the full field, avoiding masked areas, to form the random mask which is used in the LS estimator in Equation \ref{LSest}. Figure \ref{fig:rndmsk} compares the data to the random catalogs within a sample of the field created in MANGLE.

 \subsection{Measuring bias}
The linear bias in Equation \ref{bias} is used as a measure of the clustering strength of the population of quasars and has been related to many physical parameters of quasars as well as their DM environments.

Estimating the bias, however, requires that we relate the projected correlation function to the three dimensional power spectrum. To perform this task, we use Limber's approximation which projects the three-dimensional correlation function to two dimensions \citep{Limber1953} for objects with small separations ($\theta \ll 1$ rad; \citealt{Simon2007}). Projecting the correlation function requires that we integrate the three dimensional correlation function along the line of sight of two objects,
\begin{equation}\label{initproj}
\omega(\theta) = \iint \xi(r_1,r_2) \lvert r_1\rvert ^2 \lvert r_2 \rvert ^2 \phi(r_1)\phi(r_2)dr_1dr_2
\end{equation}
where $\lvert r_1\rvert, \lvert r_2 \rvert$ are the magnitudes of the two distance vectors and $\phi(r)$ is a radial selection function. The selection function acts as a probability distribution where the integral of  $r^2\phi(r)\ dr$ is normalized to unity \citep{Brewer2008}. Shifting the coordinate system to one where the unit vectors are along the line of sight, $u=r_1-r_2$, and across the line of sight, $r = \frac{1}{2}(r_1+r_2)$, Equation \eqref{initproj} can be rewritten as:
\begin{equation}\label{LimberEqn}
\omega(\theta) = \int_{0}^{\infty} r^4\phi(r)^2 dr \int_{0}^{\infty} \xi(\sqrt{u^2 + r^2\theta^2})\ du
\end{equation}
where, for small $u$, $r_1 \approx r_2$ and  for small angles, $\cos(\theta) \approx 1-\frac{\theta^2}{2}$ (see \citealt{Peebles1980}, \citealt{Brewer2008} for more details). Equation \eqref{LimberEqn} is the functional form of Limber's Equation to project the 3-D correlation function into two dimensions.

\begin{figure}[ht!]
 \centering
 \includegraphics[scale=0.45]{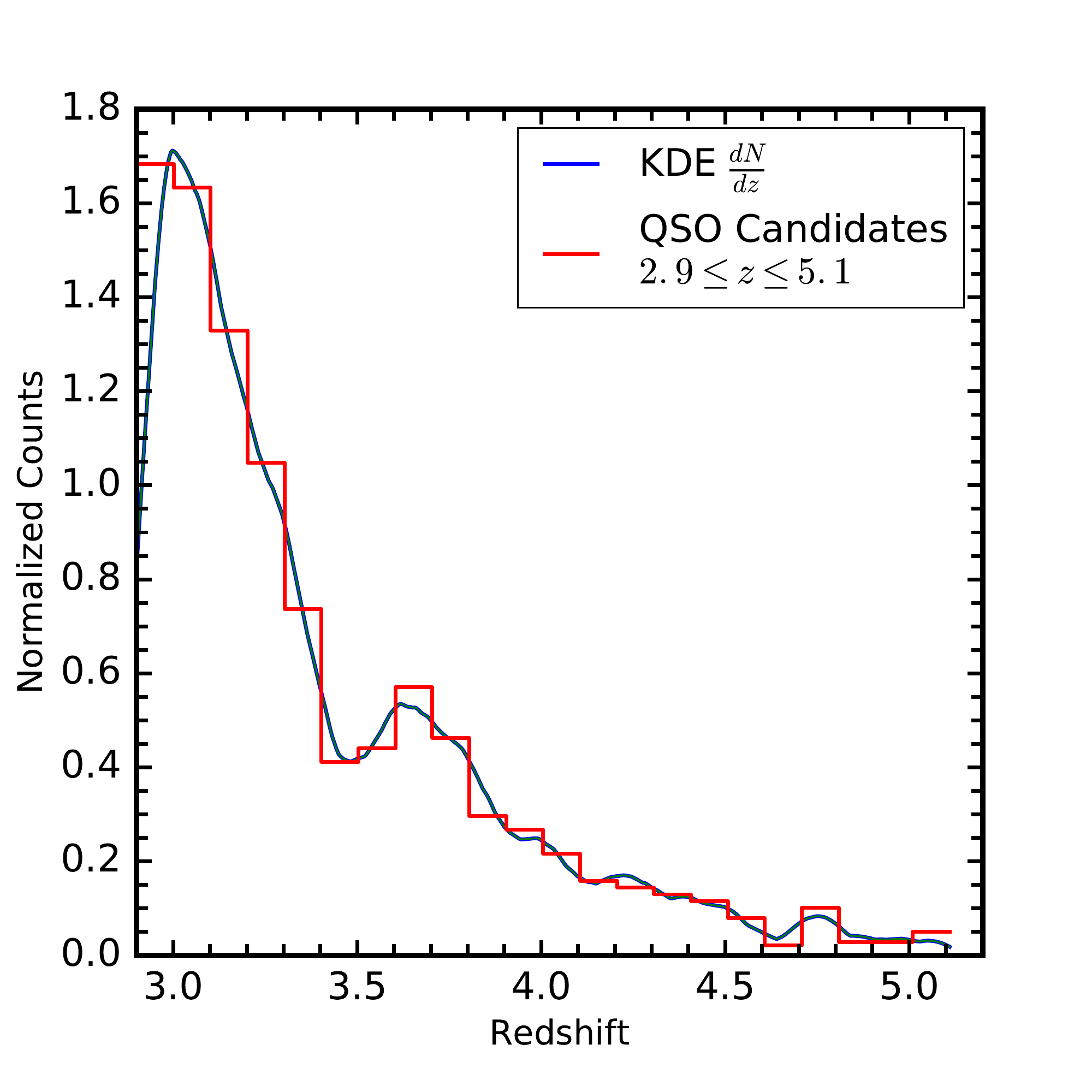}
 \caption{\footnotesize{Photometric redshift distribution of the quasar candidates. The blue curve was determined from kernel density estimation using the `epanechnikov' kernel with a bandwidth = 0.1. This curve is used in Limber's equation to estimate the bias by comparing the projected correlation function to the three dimensional dark-matter power spectrum. The red histogram depicts the distribution of the photometric redshifts in the data set.}}
 \label{dndz}
 \end{figure}

We transform Limber's equation into familiar cosmological parameters. For instance, the observed number of objects in radial shells can be described in terms of the redshift distribution of a sample of objects by:
\begin{equation}
\phi(r)r^2\ dr = \frac{dN}{dz}\ dz
\end{equation}
Solving for $\phi$ and incorporating into Limber's equation, we get:
\begin{equation}
\omega(\theta) = \int_{0}^{\infty} \left(\frac{dN}{dz}\right)^2\left(\frac{dz}{dr}\right) dz \int_{0}^{\infty} \xi(\sqrt[]{u^2 + r^2\theta^2})\ du
\end{equation}
with the variable $r$ defined as the comoving distance $\chi$ \citep{Brewer2008}. Assuming a flat Universe:
\begin{equation}
dr=d\chi = \frac{c}{H_0E_z}\ dz
\end{equation}
where $E_z =[\Omega_M(1+z)^3 + \Omega_{\Lambda}]^{\frac{1}{2}}$. Thus Equation \eqref{LimberEqn} transforms to:
\begin{equation}
\omega(\theta) = \int_{0}^{\infty} \left(\frac{dN}{dz}\right)^2\frac{H_0E_z}{c}\ dz \int_{0}^{\infty} \xi(\sqrt[]{u^2 + r^2\theta^2})\ du
\end{equation}

Using the fact that the correlation function is the Fourier transform of the power spectrum, and since we know that $u$ is small, we can employ the Hankel transformation on the second integral to obtain Limber's equation in terms of the quasar power spectrum:
\begin{equation}\label{Limbercosmo}
\omega(\theta) = \frac{H_0\pi}{c}\iint \left(\frac{dN}{dz}\right)^2E_z \frac{\Delta_Q^2(k,z)}{k^2}J_0(k\theta\chi(z))\ dk\ dz
\end{equation}
where $\Delta_Q^2$ is the dimensionless quasar power spectrum ($\Delta^2 = \frac{k^3P(k)}{2\pi^2}$) and $J_0$ is the zeroth order Bessel Function of the first kind (\citealt{Bonometto2002}; \citealt{Myers2007}; \citealt{Brewer2008}). This formula relates the 3-D quasar power spectrum to the 2-D correlation function. 

Equation \eqref{bias} can now be written in a similar fashion by replacing the correlation functions with the dimensionless power spectra of quasars and dark-matter, $\Delta_Q^2 = b^2 \Delta_{DM}^2$. We substitute this relation into Equation \eqref{Limbercosmo} which allows us to cast this equation as a function of bias directly,

\begin{equation}\label{Limberbias}
\omega(\theta) = \frac{b^2H_0\pi}{c}\iint \left(\frac{dN}{dz}\right)^2E_z \frac{\Delta_{DM}^2(k,z)}{k^2}J_0(k\theta\chi(z))\ dk\ dz
\end{equation}
where we assume that, for our samples of interest, bias does not evolve strongly with redshift or scale (e.g., \citealt{Myers2007}). Using Equation \eqref{Limberbias}, we can fit a bias value using the measurement of the projected correlation function and the 3-D dimensionless dark-matter power spectrum.

To compute the dark-matter power spectrum, we use the Code for Anisotropies in the Microwave Background (CAMB\footnote{http://camb.info}), which is a general cosmology package that creates a model cosmography. CAMB has the functionality to compute the dark-matter power spectrum including the nonlinear corrections from the halo model in \citet{Smith2003}. Combining the dark-matter power spectrum (which is a function of wave-number, $k$, and redshift, $z$) with the redshift selection function for our candidates (blue curve in Figure \ref{dndz}), we Monte Carlo integrate Equation \eqref{Limberbias} and generate a theoretical model for the projected clustering of dark-matter. Finally, we fit the DM clustering model to the measurement from our sample and obtain a bias. 
 
\section{Results}\label{sec:4}

\subsection{Projected Clustering}

\begin{figure}[h!]
\centering
\includegraphics[scale = 0.35]{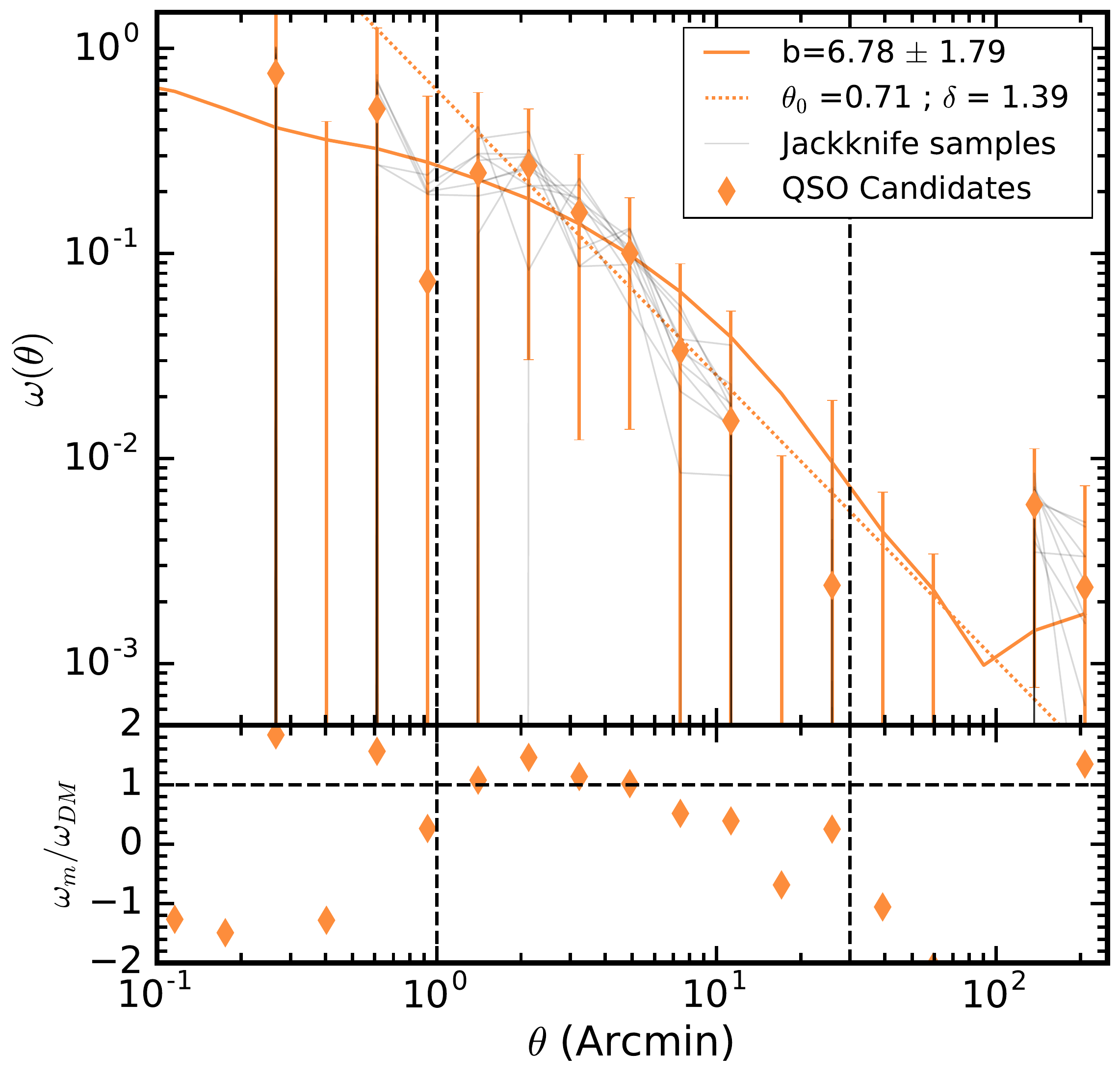}
\caption{\footnotesize{Clustering result from Table \ref{tab:Counts} of the 1378 high-redshift ($2.9 \leq z \leq 5.1$) photometric candidates (orange diamonds). Fitting the DM model to the data over the range 1$^{\prime}$ to 30$^{\prime}$ (black dashed vertical lines) produces a best fit bias of $b = 6.78 \pm 1.79$ (orange curve). This model accounts for excess power at large scales by incorporating stellar contamination into the model fit from Equation \ref{stcontam_fit}. The dotted line indicates the best fit power-law with $\theta_0$=0.71 $\pm$ 0.546 arcmin and $\delta$=1.39 $\pm$ 0.618. The lower panel shows the ratio between the measured points and the DM model. Error bars were computed using jackknife resampling, where the grey lines represent the correlation function results for each of the jackknife samples.}}
\label{fig:full_corr}
\end{figure}

The measured SpIES/SHELA angular projected 2PCF of the quasars in this sample is shown in Figure \ref{fig:full_corr}. We estimate the errors on these points using both the Poisson approximation (see Equation \ref{eqn:Poisson} in Appendix \ref{append:Errors}) along with the Jackknife resampling technique (\citealt{Scranton2002}; \citealt{Myers2007}; \citealt{Ross2009}; \citealt{Eft2015}), where a subset of the data (and the randoms) is removed from the full set, and the clustering analysis is performed on the remaining objects. In this investigation, the data sample was split into ten declination slices, resulting in ten separate clustering measurements, each excluding a different region. Using the ten jackknife clustering measurements and their RR pair counts, we compute the full covariance matrix by:
\begin{equation}
\begin{aligned}
C_{ij} = \sum_{L} \sqrt{\frac{RR_L(\theta_i)}{RR(\theta_i)}}[\omega_L(\theta_i) - \omega(\theta_i)]  \\ 
\times\ \sqrt{\frac{RR_L(\theta_j)}{RR(\theta_j)}}[\omega_L(\theta_j) - \omega(\theta_j)],
\end{aligned}\label{eqn:covariance}
\end{equation}
where $L$ denotes the {\em removal} of one of our 10 regions to form a jackknife sample comprising the other 9 regions, and $\theta_i, \theta_j$ represent the clustering result at different separation values. The error bars on the orange points in Figure \ref{fig:full_corr} show the standard deviations of the full measurement, computed by taking the square root of the main diagonal of the covariance matrix (\citealt{Myers2007}, \citealt{Ross2009}, \citealt{Eft2015}). We take Poisson errors to be the minimum error of the data, therefore we replace any Jackknife error with a value less than the Poisson estimate with the Poisson error value (see Appendix \ref{append:Errors}).

\begin{deluxetable*}{rrrrrrr}[t!]
\tablecolumns{7}
\tablewidth{0pt}
\tablecaption{Pair Counts Results}
\tablehead{
\colhead{$\theta$} & \colhead{DD}  & \colhead{DR} & \colhead{RR} & \colhead{$\omega(\theta)$} & \colhead{$\sigma_{JK}(\theta)$ }& \colhead{$\sigma_{P}(\theta)$ } \\
\colhead{(arcmin)}    & \colhead{} & \colhead{} & \colhead{} & \colhead{}& \colhead{}& \colhead{}
}
\startdata
\input{tab2}\label{tab:Counts}
\enddata
\tablecomments{Pair counts and correlation function measurements within increasing separations on the sky. Also recorded are the error estimates from the main diagonal of the covariance matrix (see Equation \ref{eqn:covariance}) estimated using jackknife resampling, as well as Poisson errors (see Equation \ref{eqn:Poisson}). In this investigation, jackknife errors are replaced with Poisson errors where the ratio of jackknife to Poisson is less than unity (see Appendix \ref{append:Errors}). In this table, we report DD and RR as double counted pairs.}
\end{deluxetable*}

The orange curve in Figure \ref{fig:full_corr}, which is a fit of the DM clustering model to the measured clustering result, incorporates an estimate of stellar contamination in the sample. Following the method in \citet{Myers2006}, stellar contamination is modeled using the measured correlation function of known stars in the field (with $g < 17.1$; \citealt{Myers2006}), as well as the efficiency, $e$, of the classification algorithms ($e=$0.86 in this study; see Table \ref{tab:classifications}). With an efficiency of $e = 0.86$, we predict 14$\%$ contamination from stellar sources in our DM model fit. The correlation function estimate becomes:
\begin{equation}\label{stcontam_fit}
\omega(\theta) = e^2 \omega_{QQ}(\theta) + (1-e^2)\omega_{SS}(\theta) + \epsilon(\theta)
\end{equation}
where $\omega_{QQ}(\theta)$ is the model result from Limber's equation, $\omega_{SS}(\theta)$ is the stellar correlation function in the field, and $\epsilon(\theta)$ is the cross correlation between quasars and stars (theoretically zero; \citealt{Myers2007}) which is insignificant in our study. Following \citet{Myers2006}, we estimate the stellar correlation function by performing the clustering analysis of SDSS point sources which have bright $g$-band magnitudes ($16.9< g <17.1$). The stellar correlation function in the footprint of this survey is $\omega_{SS} \simeq $0.1 at 30$^\prime$, slightly less than what \citet{Myers2006} found ($\omega_{SS} \simeq $0.25) using an expanded version of the KDE-selected sample of \citet{Richards2004}. We also fit a single power law to the data of the form:
\begin{equation}\label{eqn:2dpowerlaw}
\omega(\theta) = \left(\frac{\theta}{\theta_0}\right)^{-\delta},
\end{equation}
where $\theta_0$ is the angular separation over which objects are correlated, and $\delta$ defines the degree of clustering as a function of angular scale.  

Using the measurement and errors of the 2PCF (see Table \ref{tab:Counts}), and the DM model estimated using Limber's Equation, we can determine the bias that best relates the measurement and the theory. Similar to the fit in \citet{Myers2007}, the bias was fit on scales with sufficient data-data pairs ($\theta \geq 1^\prime$) and before the stellar correlation function dominates the quasar clustering signal ($\theta \leq 30^\prime$). In principle, stellar contamination does not greatly change the correlation function at small scales \citep{Myers2006}, however, photometrically-selected samples inevitably contain some level of contamination, thus it is imperative that we incorporate an estimation of  contamination in our model.

We fit the bias value, $b$, as well as the cross-correlation, $\epsilon$, over the range of $1^\prime$ to $30^\prime$ (removing the negative value points) using Equation \ref{stcontam_fit}. The best-fit bias value is $b = 6.78 \pm 1.79$ and $\epsilon = -0.010 \pm 0.018$ for the full sample of 1378 quasar candidates, which have an average redshift of $\langle z \rangle = 3.38$. Using a simple chi-squared, goodness-of-fit test $\chi^2 = 1.73$ over 5 degrees of freedom (DOF), which corresponds to a p-value of $p = 0.885$ on the fitting scales. Our model is also consistent, within error, with the data at larger scales despite fitting over the range of $1^\prime$ to $30^\prime$. This behavior reveals the effect that the stellar contaminants have and suggests that our larger-scale correlation function is contaminated with stellar sources. 

Over the same scales ($1^\prime$ to $30^\prime$), we fit the two-dimensional power-law model in Equation \ref{eqn:2dpowerlaw} to the data. The best-fit values from this two parameter model are $\theta_0 = 0.71 \pm 0.546$ and $\delta = 1.39 \pm 0.618$. Using only the best-fit amplitude of the power-law model, we estimate that the significant of this clustering result is $\sim1.3\sigma$ above the null hypothesis of an unclustered sample (i.e. $\theta_0 = 0$ at all scales). Reducing the error bars inherent to our selection technique is not practical in the near future given the depth of \emph{WISE} and the limited mapping capability of \emph{Spitzer}; however, the combination of other deep and wide-area optical and infrared data in the near future, such as The Dark Energy Survey (DES; \citealt{Diehl2014}) and \emph{Euclid} \citep{Racca2016}, should allow further progress.


\subsection{Faint Quasar Clustering}

\begin{figure}[h!]
\centering
\includegraphics[scale = 0.35]{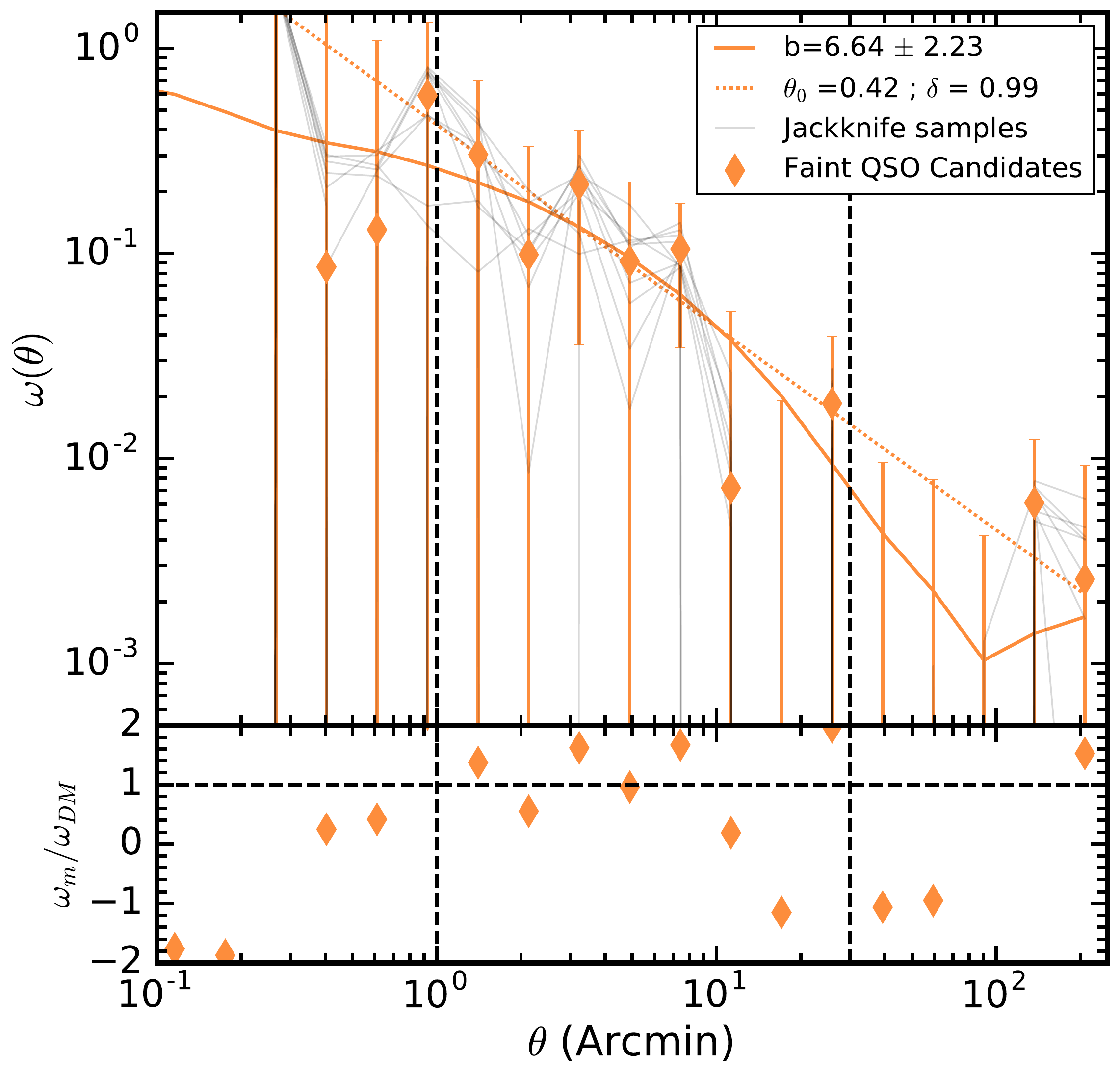}
\caption{\footnotesize{Clustering result of the faint sample of high redshift photometric candidates (orange diamonds). Fitting a new DM model to the data over the range 1$^{\prime}$ to 30$^{\prime}$ (black dashed vertical lines), we find a best fit bias of $b = 6.64 \pm 2.23$ (orange curve). Once again, we also model stellar contamination using Equation \ref{stcontam_fit} with the new selection function for the faint objects. The dotted line indicates the best fit power-law with $\theta_0$=0.42 $\pm$ 0.582 arcmin and $\delta$=0.99 $\pm$ 0.502. As in Figure \ref{fig:full_corr}, we show the ratio of the data to the DM model in the lower panel and errors are computed with jackknife resampling (grey lines).}}
\label{fig:faint_corr}
\end{figure}

The results in Figure \ref{fig:full_corr} show the clustering strength of \emph{all} of our candidate quasars, both bright ($i < 20.2$; 252 objects) and faint ($i \geq 20.2$; 1126 objects). In this analysis, we remove the bright quasars and cluster \emph{only} the 1126 faint objects to directly test the degeneracy in the models of \citet{Hopkins2007}. The computation of the correlation function and bias is the same as the previous section, we simply change the redshift selection function in Limber's equation to match the new distribution. We find a best fit bias of $b = 6.64 \pm 2.23$ and $\epsilon = 0.005 \pm 0.022$ for this faint sample with an average redshift of $\langle z \rangle = 3.39$. The chi-squared test results in $\chi^2 = 0.45$, again over 5 degrees of freedom (DOF), which corresponds to a p-value of $p = 0.994$ on the fitting scales. The results of this analysis are shown in Figure \ref{fig:faint_corr}. The error in this fit is much larger than in the full sample which we attribute to the size of the error bar at $\sim$ 5 arcmin and the difference in value at $\sim$ 30 arcmin, which are both likely due to a smaller number density of objects. Despite this difference, the bias between this sample and the full sample are consistent; however, we focus on the full sample results in the next section.

\section{Implications}\label{sec:5}
\subsection{Comparison to Other Observations}\label{sec:comparisons}

This paper presents the first measurements of the autocorrelation function of photometrically-selected high-$z$ quasars; however, there are other measurements of quasar clustering with which we can compare across our redshift range of interest. Here we examine the techniques and results of the surveys in the literature to those in our study. 

\begin{figure*}[ht!]
\centering
\includegraphics[scale = 0.80]{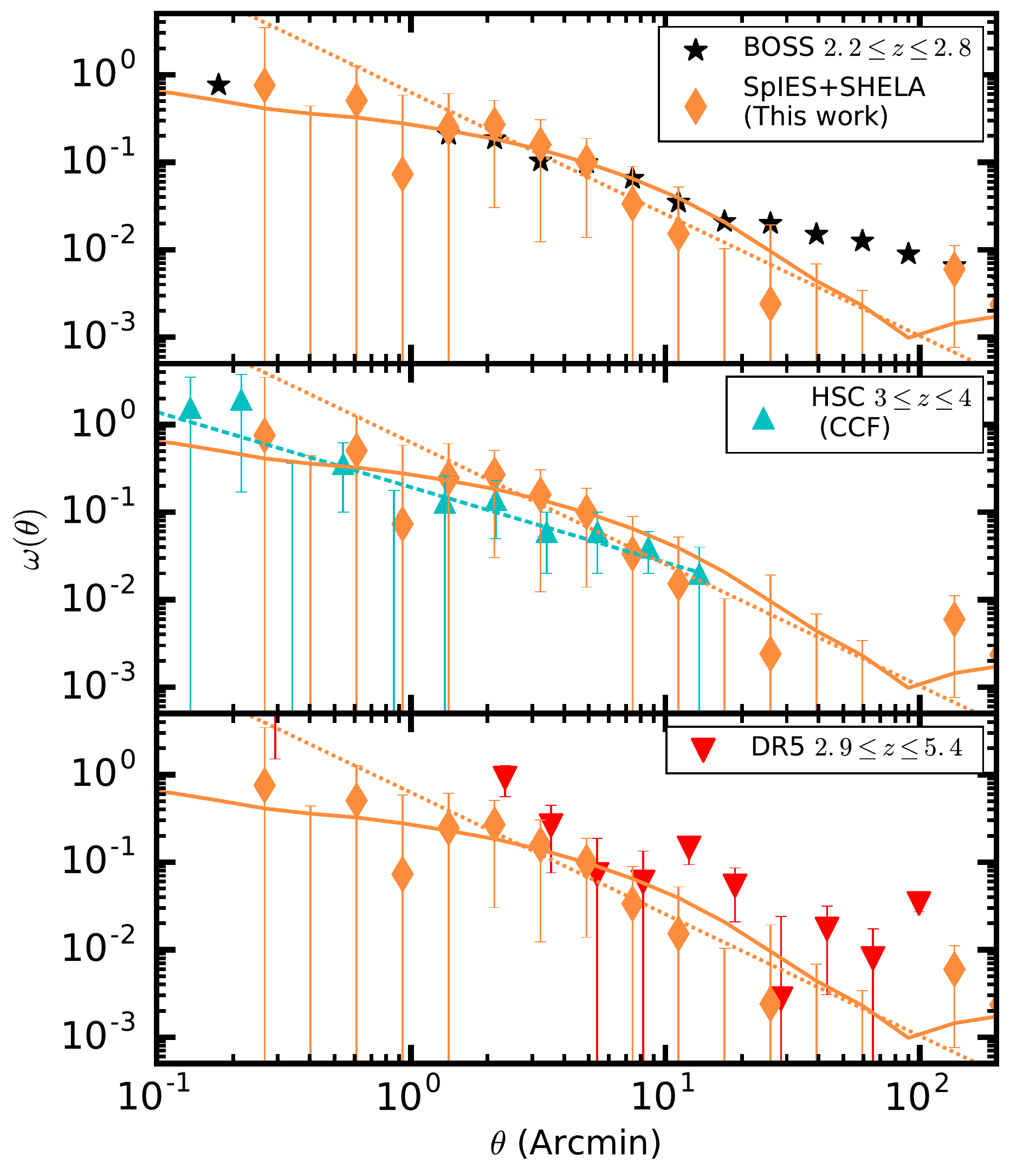}
\caption{\footnotesize{Top: Comparison of the clustering measurement from this study (orange diamonds) to the angular correlation function from the BOSS survey (black stars), which was computed using a subset of the data from \citet{Eft2015}. Middle: Comparison to the CCF results of \citet{He2017} (light blue triangles). While the ACF (our study) and CCF \citep{He2017} cannot be directly compared, these results cover approximately the same redshift range, and have slightly different bias values (see Figure \ref{fig:BZ}). Bottom: Clustering results of the full redshift range in this study compared to that of spectroscopically-confirmed quasars from SDSS Data Release Five (inverted red triangles; \citealt{Shen2007}). The two surveys cover the same redshift range; however, the quasars in this study are significantly fainter than those in \citet{Shen2007}, as shown in Figure \ref{fig:Abs_mag}. Poisson error bars are depicted for the \citet{Shen2007} data, using the data-data pair counts we estimate in our analysis. Points for \citet{Shen2007} are offset by 0.1$\times \theta$ arcmin for clarity. We compare with these three surveys since they are closest in redshift range (although not exactly the same), and are consistent on scales before contamination dominates ($\sim 20^\prime$; \citealt{Eft2015}).}}
\label{fig:cor_compare}
\end{figure*}

\begin{deluxetable*}{lccrccccc}[h!]
\tablecolumns{9}
\tablewidth{0pt}
\tablecaption{High-$z$ Bias Measurements}
\tablehead{
\colhead{Measurement} & \colhead{$z$-interval} & \colhead{$\langle z \rangle$} & \colhead{$N_{qso}$} & \colhead{bias} & \colhead{$\theta_0$} & \colhead{$\delta$} & \colhead{M$_i$[z=2]}& \colhead{M$_{DMH}$} \\
\colhead{} &\colhead{} & \colhead{} & \colhead{} & \colhead{} & \colhead{(arcmin)} & \colhead{} & \colhead{faint, bright}& \colhead{($\times10^{12}h^{-1}M_{\odot}$)}
}
\startdata
This work (all) & 2.90, 5.10 & 3.48& 1 378 & 6.78 $\pm$ 1.79 & 0.710 $\pm$ 0.546 & 1.39 $\pm$ 0.618 & -23.80, -27.50 & 1.70 -- 9.83 \\
This work (faint) & 2.90, 5.10 & 3.49& 1 126 & 6.64 $\pm$ 2.23 & 0.420 $\pm$ 0.582 & 0.99 $\pm$ 0.502 & -23.80, -26.40& 1.04 -- 10.6  \\
\citet{He2017}$^{a}$  &3.00, 4.00 & 3.80 & 901 & 5.93 $\pm$ 1.43 & 0.148 $\pm$ 0.050 & 0.86$^{a}$ & -23.70, -25.86& 1.00 -- 2.00 \\
\citet{Eft2015}$^{b}$ & 2.64, 3.40 & 2.97 & 24 724 & 3.57 $\pm$ 0.09 & -- & -- & -24.40, -29.31& 0.60 -- 0.72  \\
\citet{Shen2007}$^{b,c}$  & 2.90, 3.50 & 3.20& 2 651 & 7.90 $\pm$ 0.80 & -- & -- & -26.00, -30.00& 2.00 -- 3.00 \\
\citet{Shen2007}$^{b,c}$  & 3.50, 5.40 & 4.00& 1 775 & 14.0 $\pm$ 2.00 & -- & -- & -26.50, -30.00& 4.00 -- 6.00

\enddata
\tablecomments{Bias estimates for selected surveys of comparable redshifts to our study. \\
$^{a}$ Cross-correlation of the faint sample. Power law index held fixed at $\delta = 0.86$ in this study.\\
$^{b}$ Redshift space estimate, thus no angular power law information is given.\\
$^{c}$ \citet{Shen2007} results split into two redshift bins to reflect the bias values shown in Figure \ref{fig:BZ}.
}\label{tab:bz}
\end{deluxetable*}

We first compare our results to the results of the BOSS survey from \citet{Eft2015}. This study examined the redshift-space correlation function of spectroscopically-confirmed quasars in the SDSS field in the redshift range of 2.2 $\leq z \leq$ 2.8. For a more direct comparison with our angular-projected correlation function, we compute the angular correlation function of the BOSS data from \citet{Eft2015}, using their NGC CORE sample and a random catalog with five times the data; shown in the top panel of Figure \ref{fig:cor_compare}. Despite spanning slightly disjoint redshift ranges, the two correlation functions agree on scales before contamination dominates the signal ($\sim 25 h^{-1}$ Mpc or $\sim 20^\prime$; \citealt{Eft2015}). Since these correlation functions have similar power in the clustering signal, yet are at different redshifts, the best fit bias values are different (see Figure \ref{fig:BZ}). 

Next, we compare with the results of \citet{He2017}, who computed the quasar cross-correlation function (as opposed our measurement of the auto correlation function; ACF) for photometrically selected quasars in the redshift range 3 $\leq z \leq$ 4. In the \citet{He2017} investigation, quasars are selected using optical and near-infrared colors from the Hyper Suprime-Cam\footnote{\url{https://www.naoj.org/Projects/HSC/surveyplan.html}} (HSC). In total, they selected 1023 quasars as candidates across 172 deg$^2$, 901 of which were both faint ($i \geq$ 21) and high-$z$. Using these candidates, they computed the cross-correlation function (CCF) between their candidates and Lyman-Break Galaxies at $z \sim$ 4. Figure \ref{fig:cor_compare} (middle panel) depicts the results from the CCF analysis compared to our study. Since the measurement is performed with two different statistics, the amplitudes of the two $w(\theta)$ should not be directly compared; however the bias measurements from these two surveys can be compared, despite being computed with different statistics (CCF and ACF). Our ACF measurements find a bias of $b$ = 6.78 $\pm$ 1.79, and the bias from the CCF of the less-luminous ($i \geq$ 21) quasars in \citet{He2017} is $b$ = 5.93 $\pm$ 1.43; both results are displayed in Figure \ref{fig:BZ}. The biases of these two studies overlap within their measurement error, and can be interpreted using a similar physical model. We will discuss the physical implications of this model in Section \ref{sec:feedbackmod}. A larger sample of spectroscopic high-$z$ quasars is needed to reduce the uncertainties in the bias measurement of high-$z$ quasars. 

We also compare our study over the full redshift range to the results of \citet{Shen2007}, who investigated the clustering properties of spectroscopically-confirmed high-$z$ quasars from SDSS Data Release five (DR5). These DR5 quasars span a redshift range of 2.9 $\leq z \leq$ 5.4, and are bright ($i\leq$ 20.2; see Figure \ref{fig:Abs_mag}). With spectroscopic redshifts, \citet{Shen2007} present a measurement of the 3D redshift-space correlation function, so to compare their results to ours, we compute the angular projected correlation function using their data and the DR5 mask from \citet{Ross2009}. The results are shown in the bottom panel of Figure \ref{fig:cor_compare}. The correlation function is, in general, higher in amplitude for the objects in DR5 than our candidates over the relevant scales ($\sim 30^\prime$), however we find a slightly smaller bias value than \citet{Shen2007}.  

The \citet{Shen2007} quasar sample has an i-band limiting magnitude of M$_{i} = -26.5$ (their Table 6), and is thus only sampling the very bright end of the quasar luminosity function. By contrast, our data as well as the data from \citet{He2017} have an i-band limiting magnitude of M$_{i} \simeq -24.0$. A direct comparison of the bias values (see Figure \ref{fig:BZ}) between \citet{Shen2007}, \citet{He2017}, and our study hints at a level of luminosity-dependence of clustering for high-$z$ ($z \geq 3$) quasars. This difference in clustering would suggest that, at $z \geq$ 3, the mass of the dark-matter (DM) halo hosting bright quasars is larger than the host DM halo masses of low luminosity quasars. Luminosity dependence at high-$z$ would be a fascinating result since, at low-$z$, it has been shown that clustering is weakly dependent on luminosity, if at all (\citealt{daAngela2008}; \citealt{Shen2009}; \citealt{Eft2015}; \citealt{Chehade2016}).

\subsection{Dark Matter Halo Mass}
Using the measured quasar bias in this study, and the hypothesis that quasars are biased tracers of the underlying DM distribution, we can estimate the characteristic mass for a typical DM halo. Here we use the formalism of \citet{Tinker2010}, who fit analytic models to the results of simulated clustering of DM halos in a flat $\Lambda$CDM cosmology. We adopt the fitting function in Equation (6) of \citet{Tinker2010}:

\begin{equation}\label{eqn:dmhalofit}
b(\nu) = 1 - A\frac{\nu^a}{\nu^a + \delta_c^a} + B\nu^b+C\nu^c,
\end{equation}
where $b(\nu)$ is the measured bias in our study and $\nu$ is the ``peak height'' of the density field defined by $\nu = \delta_c/\sigma(M)$. Here, the peak height is defined in terms of the critical density for collapse of the DM halo ($\delta_c$=1.686) and the linear matter variance at the radial scale of each halo, $R_{halo} = (3M_{halo}/4\pi\bar{\rho}_m)^{1/3}$ ($\bar{\rho}_m=2.78\times 10^{11}\Omega_mh^2M_{\odot}$; \citealt{He2017}), defined by:
\begin{equation}
\sigma^2(M) = \frac{1}{2\pi^2}\int P(k,z)\hat{W}^2(k,R)k^2dk.
\end{equation}
We estimate the matter power spectrum, $P(k,z)$, using CAMB and our adopted cosmology, where $\hat{W}(k,R)$ is the spherical top-hat window function;
\begin{equation}
\hat{W}(k,R) = \frac{3}{(kR)^3}\left(\mathrm{sin}(kR)-kR\mathrm{cos}(kR)\right).
\end{equation}
The parameters $A$, $a$, $B$, $b$, $C$, $c$  in Equation \eqref{eqn:dmhalofit} are adopted from Table 2 of \citet{Tinker2010} for $\Delta = 200$, where $\Delta$ is the ratio of mean density to background density (similarly used in \citealt{Eft2015}, \citealt{DiPompeo2016}, \citealt{He2017}):
\begin{gather}
y = \mathrm{log}_{10}(\Delta) \nonumber \\
A = 1+0.24ye^{-(4/y)^4} \nonumber \\
a = 0.44y-0.88 \nonumber\\
B = 0.183 \\ 
b = 1.5 \nonumber\\
C = 0.019+0.107y+0.19e^{-(4/y)^4} \nonumber\\
c = 2.4 \nonumber
\label{eqn:params}
\end{gather}

Using the measured bias values in Equation \ref{eqn:dmhalofit}, the power spectrum from CAMB, and the parameters defined above, we can solve for the characteristic halo mass (see Table \ref{tab:bz}). For our measured bias over the full redshift range of $b = 6.78 \pm 1.79$, the characteristic halo mass ranges between 1.70--9.83$\times 10^{12} h^{-1}M_{\odot}$. Computing the halo mass from the bias estimated using only the faint quasars, yields 1.04--10.56$\times 10^{12} h^{-1}M_{\odot}$, where the large mass ranges in both estimates are a direct result of the large uncertainty in the bias values.

We compare our estimated halo masses to the masses found in \citet{Shen2007} who computed the minimum halo mass, which is slightly different from our computation in that an estimate of the luminosity function is required. Over the redshift range of $2.9 \leq z \leq 3.5$, \citet{Shen2007} find a minimum halo mass of $\sim$(2-3)$\times 10^{12} h^{-1}M_{\odot}$, and in the redshift range $z \geq 3.5$, \citet{Shen2007} estimates a minimum halo mass of $\sim$(4--6)$\times 10^{12} h^{-1}M_{\odot}$. 

The low-$z$ halo mass estimate from \citet{Eft2015} of $\sim$0.66 $\times 10^{12} h^{-1}M_{\odot}$ over the redshift range of $2.64 \leq z \leq 3.4$ (their Table 7), is a factor of ten smaller than our results; however they also report halo masses on the redshift range $2.20 \leq z \leq 2.80$ of $\sim$ 1.2--2.8 $\times 10^{12} h^{-1}M_{\odot}$, which is $\sim$ 3$\times$ smaller than our result. This difference arises from the different redshifts as well as the large difference in bias. The high-$z$ estimate of the \citet{He2017} less-luminous sample is 1--2$\times 10^{12} h^{-1}M_{\odot}$. Again, the difference here is mainly due to the difference in bias between the two studies. However, if we take these results at face value it does imply that less luminous quasars tend to have smaller halo masses at high-$z$. A larger sample of spectroscopically confirmed faint, high-$z$ quasars is needed to answer this question with greater certainty. If we could increase the number of pair counts along the fitting scales by 50$\%$, we estimate that the error bars would decrease by $\sim$ 20$\%$ (using the Poisson error estimate which scales as $DD^{-0.5}$). More data would reduce the error on the bias which, in turn, leads to a tighter constraint on the DM halo masses.

\subsection{Implications for Feedback}\label{sec:feedbackmod}

\begin{figure}[b!]
\includegraphics[scale = 0.35]{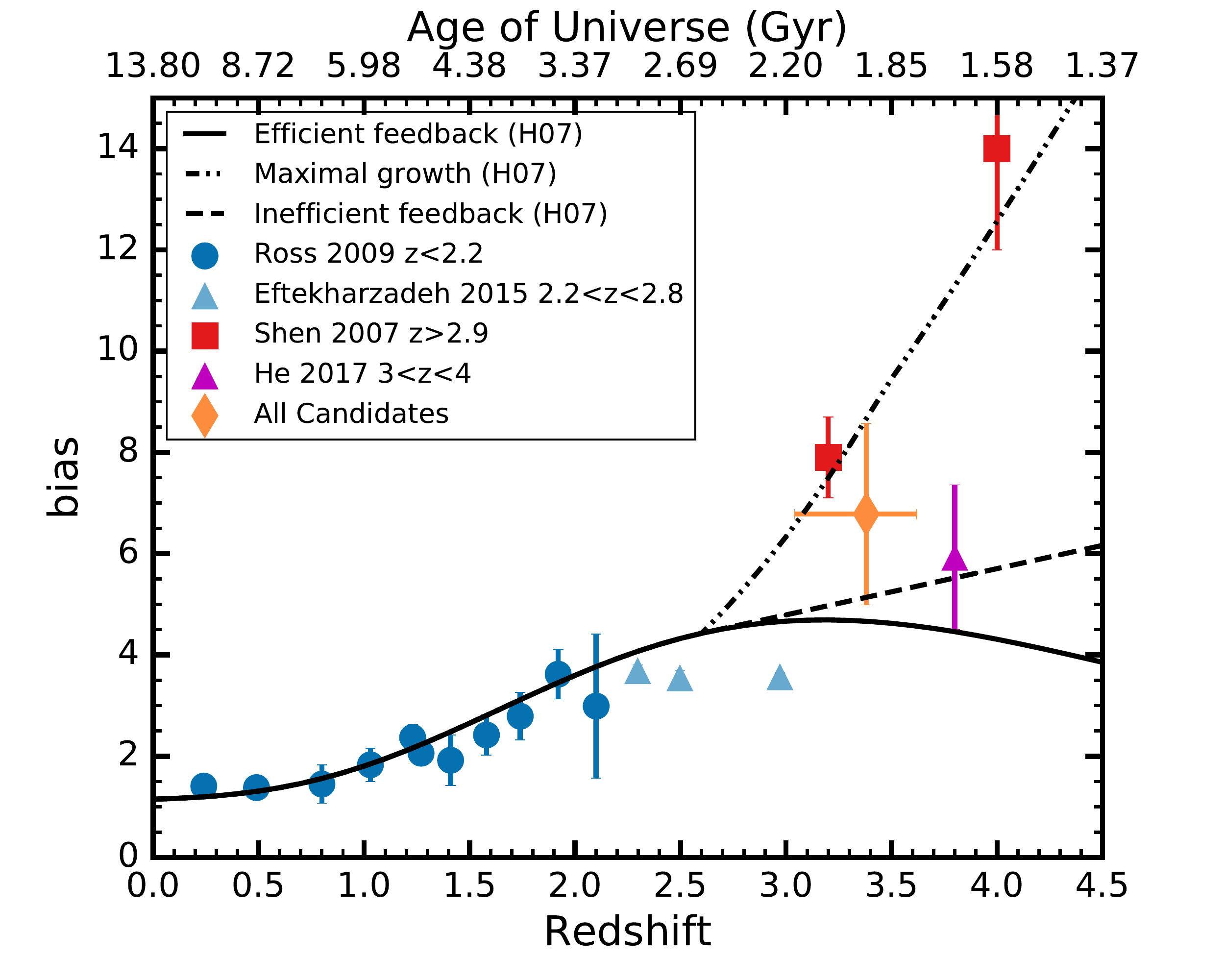}
\centering
\caption{\footnotesize{The evolution of the bias with redshift. We show the bias result for our full candidate sample (orange diamond). Also displayed are the feedback models from \citet{Hopkins2007} as well as the low- and intermediate redshift measurements from \citet{Ross2009} (dark blue circles) and \citet{Eft2015} (light blue triangles), respectively. Finally, we show the high-$z$ bias of the bright quasars from \citet{Shen2007} (red squares) and the new HSC study from \citet{He2017} (purple triangle). The bias increases with redshift in our sample and tends to agree with the ``inefficient feedback'' model, however we cannot rule out the ``maximal growth'' model.}}
\label{fig:BZ}
\end{figure}

The measurement of the 2PCF and bias of the faint, high-$z$ quasars in this study is ideal information to constrain the feedback mechanisms presented in \citet{Hopkins2007}. The \citet{Hopkins2007} study compared the clustering of quasars and galaxies as a function of different intrinsic properties (e.g.,\ mass, luminosity, redshift) to investigate triggering mechanisms and the growth of the quasar and galaxy populations. Included in this study was an analysis of how different quasar feedback mechanisms affect their clustering strength. These models were designed to fit measured results at low-$z$ (e.g.,\ \citealt{Croom2005}, among others) which we represent using the \citet{Ross2009} results, yet vary at high-$z$ ($z \geq 3$). This study highlighted three feedback scenarios; {\it{``efficient''}} and {\it{``inefficient''}} feedback, as well as a {\it{``maximal growth''}} model. We depict the clustering predictions from these three models as the black lines in Figure \ref{fig:BZ}, and provide a brief explanation below.

The solid line in Figure \ref{fig:BZ} depicts the clustering evolution with redshift if BH growth shuts down after the quasar epoch. This is the ``efficient'' feedback model in \citet{Hopkins2007}, and assumes that quasars represent a single, short-lived, phase in the growth of the central BH. Here, feedback efficiently terminates the quasar phase, and the central BH ceases its growth. This model assumes that the observed properties of quasars at $z < 2$ are the same as at higher redshifts, thus the predicted clustering strength weakens at high-$z$ to reflect observations at low-$z$.

Additionally, \citet{Hopkins2007} presents a model in which quasars, and their central BHs, grow intermittently until $z \sim 2.5$ when ``downsizing'' begins (the dashed line in Figure \ref{fig:BZ}). In this model, the quasar grows with the luminosity function, and the evolution of the luminosity function is dictated by the same objects growing hierarchically. Thus, feedback is ``inefficient'' since the BH continues to grow over various epochs, as opposed to the first model where, after the initial quasar phase, BH growth ends. This also means that brighter quasars live in very massive DM halos and fainter quasars would live in smaller DM halos at early times.

The ``maximal growth'' model postulates that the central BHs continue to grow proportionally with the DM halo until $z \sim 2$. This model assumes that quasars are continually accreting at their Eddington rates. Here, feedback is not only inefficient prior to $z \sim 2$ but is not sufficient to stop the BH from growing at its most maximal rate. These quasars live in the highest mass DM halos which accumulates gas unimpeded by the radiation from the central quasar. Therefore, the predicted clustering is very high from this model is shown by the dot-dashed line in Figure \ref{fig:BZ}. 

At low-$z$, the three models are designed to match measurements of the 2PCF (\citealt{Croom2005}; \citealt{Myers2007}; \citealt{Ross2009}), but beyond $z \sim 3$, the models diverge. Additionally, the three models become degenerate for a sample of quasars with $i \leq 20.2$ \citep{Hopkins2007}; all taking the form of the ``maximal growth'' model. Our study, however, examines quasars fainter than the limit at high-$z$, thus breaking the degeneracy between the models in the redshift range $3 \leq z \leq 4$. 

Figure \ref{fig:BZ} displays the best fit bias result over all of our candidates over the full redshift range in this study (orange diamond). The bias of the of faint candidates ($i \geq$ 20.2) is not depicted, however is consistent with the full result. Also depicted in Figure \ref{fig:BZ} is the error in both the bias, which is a result from fitting the dark-matter model, and in redshift, where, since our redshift distribution is not Gaussian, we depict the first and third quartile of the redshifts (as opposed to the standard deviation). 

Within the error of these results, the bias in our study overlaps both the ``maximal growth'' model and the ``inefficient feedback'' model, as shown in Figure \ref{fig:BZ}, for the full sample of candidates in this analysis. The ``maximal growth'' model is also consistent with the results of \citet{Shen2007}; however, we remind the reader that our investigation clustered a different population of quasars than \citet{Shen2007}. We analyzed the clustering of \emph{faint} quasars and are therefore capable of breaking the degeneracy limit noted in \citet{Hopkins2007}. As shown in Figure \ref{fig:BZ}, our result deviates from the ``maximal growth'' model toward the ``inefficient feedback'' model, which coincides with the result from \citet{He2017} at $z \sim 4$. The ``inefficient feedback'' model predicts that feedback from the central BH intermittently shuts down the accretion of gas onto the BH at early times. This model also suggests a degree of luminosity dependence of quasar clustering at high-$z$ and that fainter quasars live in less massive DM halos as compared to bright quasars. To better understand these models at $z \sim 3.4$ will likely require a larger sample of spectroscopically-confirmed quasars that are both faint, and high-redshift.

At first glance, it may appear that the findings in \citet{Eft2015} contradict our results; however, a significant difference in the bias measurements between our study and \citet{Eft2015} can be attributed to the difference in the redshift selection functions. While Figure \ref{fig:cor_compare} shows that our results and the angular correlation function of \citet{Eft2015} have a similar amplitude, the DM model is strongly dependent on the redshift selection function. Lower redshift ranges result in larger power in the angular correlation function model which, in turn, results in a smaller bias fit (i.e.,\ decreasing redshift in the model shifts the orange curve in Figure \ref{fig:cor_compare} to the right). As a result, we expect \citet{Eft2015} to have a lower bias than our investigation despite having similar amplitudes in angular correlation space. Taking these bias values at face value shows a rapid change in the bias at $z \sim$ 3.1. Understanding this jump in bias at this particular redshift will be the topic of future work.

\section{Summary}\label{sec:6}
In this investigation, we have determined the two-point autocorrelation function of 1378 photometrically-selected, faint ($i \geq 20.2$), high-$z$ ($2.9 \leq z \leq 5.1$) quasars across $\sim$100 deg$^2$ on SDSS S82. Details about this catalog as well as our main findings are as follows:

\begin{itemize}
    \item{We combine the deep optical photometry on S82 from SDSS with new, deep MIR information from the SpIES and SHELA surveys to form a comprehensive catalog of photometric objects. Utilizing their optical/MIR colors, and the colors of known high-$z$ quasars from the \citet{Richards2015} composite catalog (see Figure \ref{fig:traincol}), we use three machine-learning algorithms to select 1378 faint, high-$z$ quasar candidates.}

    \item{We estimate the photometric redshifts of these candidates using Nadaraya-Watson kernel regression. When tested on spectroscopic quasars, this algorithm predicts photometric redshifts within a range of $z_{phot}-z_{spec} = 0.1$ for 93\% of the quasars (Figure \ref{fig:photvspec}). The overlap in color-redshift space between the photometric candidates and the known quasars with which they were selected is presented (Figure \ref{fig:colzspecsing}).}

    \item{Figure \ref{fig:Abs_mag} demonstrates that our candidates are generally fainter than the objects used in the \citet{Shen2007} study. This aspect of our sample helps to break the degeneracy between the feedback models studied in \citet{Hopkins2007}.}

    \item{Utilizing the estimator from \citet{Landy1993}, we compute the angular 2PCF of our faint high-$z$ quasars, where a random mask is generated using MANGLE (Figure \ref{fig:rndmsk}). The correlation function result is presented in Figure \ref{fig:full_corr}}
    
    \item{We estimate a linear bias using the method of \citet{Limber1953} which relates the 3D DM power spectrum to the angular correlation function. We compute the 3D power spectrum using CAMB and our fiducial cosmology. Over the full redshift range of our sample ($\langle z \rangle=3.38$), the bias is $b=6.78 \pm 1.79$. The best-fit values from the power law model are $\theta_0 = 0.71 \pm 0.546$ and $\delta = 1.39 \pm 0.618$.}
    
    \item{In Figure \ref{fig:faint_corr}, we remove the bright objects and recompute the correlation function of 1126 faint quasar candidates. We find the faint quasars have a bias of $b=6.64 \pm 2.23$, similar to the full study. The agreement in bias demonstrates that the bright quasars in the sample do not skew the bias result of the faint objects. We compare the results of our full study with other surveys in Figure \ref{fig:cor_compare}.}
    
    \item{Using the estimates of bias, we compute characteristic DM halo masses using the formalism of \citet{Tinker2010}. Our quasars inhabit DM halos with masses of 1.70--9.83$\times 10^{12}h^{-1} M_{\odot}$. This mass estimate covers a wide range due to the large uncertainty in the bias.}
    
    \item{We use our bias estimate to constrain the feedback models of \citet{Hopkins2007} in Figure \ref{fig:BZ}. Our data is consistent with both the ``maximal growth'' model, which assumes that the central quasar is not powerful enough to shut down accretion of material onto the BH, as well as the ``inefficient feedback'' model, which suggests that feedback from the central source intermittently shuts down accretion of the central BH. The ``inefficient feedback'' model, however, also coincides with the bias of faint quasars at $z \sim 4$ found in \citet{He2017}. Finally, the ``inefficient feedback'' model suggests that fainter quasars sit in smaller DM halos.}

\end{itemize}

Further studies of the 2PCF of faint, high-$z$ quasars will benefit from the new optical and infrared surveys on the horizon. Surveys performed with the Large Synoptic Survey Telescope (LSST; \citealt{Abell2009}) in the optical and the Wide-Field Infrared Survey Telescope (WFIRST; \citealt{Spergel2013}) and, to an extent, the James Webb Space Telescope (JWST; \citealt{Gardner2006}) in the infrared will be able to observe fainter than what we have now. These surveys will add an immense amount of data to our sample and a significant amount of area which, in turn, increases the significance of the results. Similarly, spectroscopic investigation on the candidates will allow us to add to the high-$z$ training data, as well as make the necessary corrections to our photometric redshifts to compute the redshift-space 2PCF. In this investigation, however, we have demonstrated that, using machine-learning techniques, we can both select faint, high-$z$ quasars cleanly and compute the 2PCF on these samples. These techniques will be crucial in the next phase of astronomy, which will be dominated by photometric data that lacks detailed spectroscopic follow-up.

\subsection*{\normalfont{Acknowledgements}}
This work is based on observations made with the \emph{Spitzer} Space Telescope, which is operated by the Jet Propulsion Laboratory, California Institute of Technology under a contract with NASA. Support for this work was provided by NASA through an award issued by JPL/Caltech. We acknowledge support from NASA-ADAP grant NNX17AF04G. ADM was partially supported by NSF grants 1515404 and 1616168 and NASA grant NNX16AN48G. FEB acknowledges support from CONICYT-Chile 
(Basal-CATA PFB-06/2007, FONDECYT Regular 1141218), the Ministry of Economy, Development, and Tourism's Millennium Science Initiative through grant IC120009, awarded to The Millennium Institute of Astrophysics, MAS. GTR acknowledges support from NSF grant 1411773.

We make our full datasets, analyses code and methodologies available at
\url{https://github.com/JDTimlin/QSO_Clustering/tree/master/highz_clustering}

For this research, we use the Python language and \href{http://www.astropy.org/}{Astropy}\footnote{astropy.org} \citep{astropy}. \href{http://www.star.bristol.ac.uk/~mbt/topcat/}{TOPCAT}\footnote{starlink.ac.uk/topcat} \citep{TOPCAT}. 

We thank Michael DiPompeo and Ryan Hickox for their correspondence and advice on both our contamination checks and in our interpretations of the feedback models. We also thank Yao-Yuan Mao for the DECaLS image list tool (\url{https://github.com/yymao/decals-image-list-tool}) used in this study.

\appendix
\section{Contamination Checks}\label{append:contamination}
Contamination in any clustering sample can drastically change the correlation function and the resulting bias. We carefully define our sample in this study to avoid contamination as much as possible. As part of this work, we also performed a clustering analysis using the selection results without restricting to the point sources alone. We found that, if we just use color selection and do not check for low-$z$ contamination, we get a bias value of $b \sim$ 5 instead of $b \sim$ 6.5, which would lead us to different conclusions in Figure \ref{fig:BZ}. It is therefore very important that we eliminate as much contamination as possible in this study. 

While we explicitly model stellar contamination in this study, there are other forms of contamination that dilute the clustering signal. The two main sources of additional contamination are mis-identification of objects in the classification algorithms, and regions where the angular mask of the random objects is not identical to the data. Here we describe our methods to identify and reduce contamination from galaxies in our analysis.

\subsection{Extinction Cut}
\begin{figure}[h!]
\includegraphics[scale = 0.5]{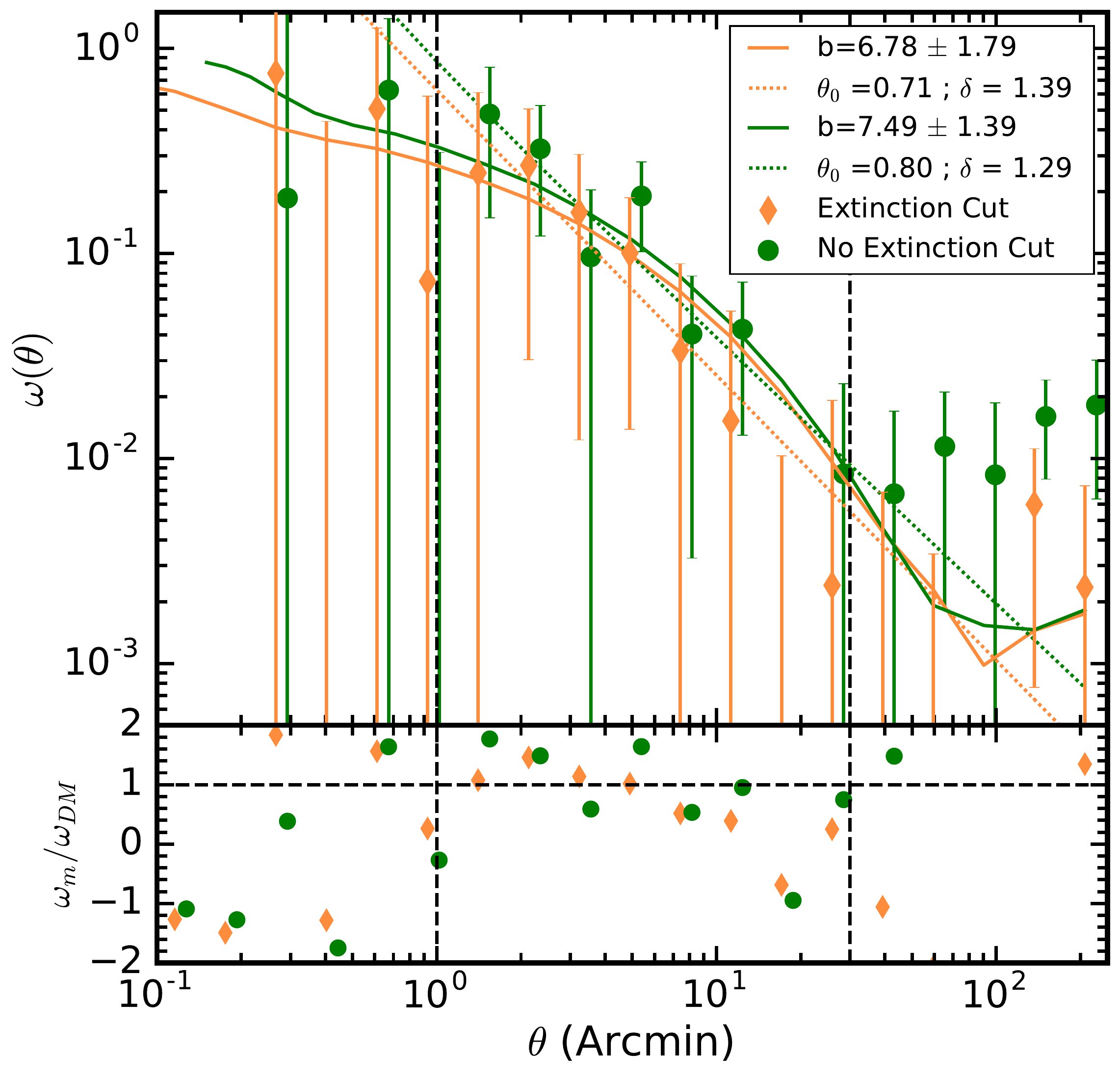}
\centering
\caption{\footnotesize{Correlation function of the final sample after the extinction cut (orange diamonds) compared to the correlation function of the full sample of objects, including the region $330 \leq \alpha_{J2000} \leq 344.4$ (green circles; offset by 0.1 arcmin for clarity). While the correlation function is similar over the fitting range ($1^\prime \leq \theta \leq 30^\prime$), the power at larger scale is significantly higher for the full study compared to the extinction cut survey. The green circles are offset by $0.1*\theta$ for clarity.}}\label{fig:extcut}
\end{figure}

As mentioned in Section \ref{sec:clustering_sample}, we cut the overlapping region between SpIES and the outskirts of the disk of the Milky Way ($330 \leq \alpha_{J2000} \leq 344.4$ which corresponds to a galactic latitude of $-51.5 \leq b_{gal} \leq -41.5$) to eliminate highly-extincted objects from the analysis which act as contaminants in the clustering signal. Figure \ref{fig:extcut} depicts the clustering result before (green circles) and after (orange diamonds) this extinction cut, as well as their best fit DM models (which have slightly different redshift selection functions). These models are fit as before using an efficiency of $e = 0.86$ which means that 14$\%$ of the sample are stellar contaminants. At large scales, the model (green curve) lies below the measured clustering strength, which implies that there are more contaminants than estimated using just stellar contamination. After the extinction cut is performed, however, there is much better agreement between the model and the data (in fact, it appears that the model over-estimates the contamination at large scales). Deep infrared spectra are required to determine the particular type of object contaminating the sample, however it is most likely stars that were reddened by Galactic dust such as late type M-dwarf stars. These objects would not appear in optically--selected samples, however since we include the infrared colors in our selection, they could be selected as quasars. 

While the extinction cut resulted in a loss of $\sim$ 20 deg$^{2}$, it also significantly decreased the power of the correlation function at larger scales (see Figure \ref{fig:extcut}). There were, however, objects in that field with lower extinction measurements that were also cut. Ideally, we would keep these objects to use in our correlation function measurements, but cutting on the extinction value causes the density to drop significantly in this area, which affects the correlation function if not properly accounted for in the angular mask. Our future work to remedy this problem is to change the density of the random mask in this field to reflect the data.

\subsection{Visual Inspection}
Visual inspection using the DECam Legacy Survey image cutout tool\footnote{\url{https://github.com/yymao/decals-image-list-tool}} enabled us to examine the classified objects and eliminate obvious sources of contamination. The superior depth and resolution of DECam is crucial for the follow-up visual inspection of the candidates that were selected in each algorithm. This inspection also drove the need to create the point-source metric we used to cut all extended objects in this study. We note that fainter quasars are more likely to be classified as extended emission, thus spectroscopic follow-up is needed on \emph{all} faint candidates, not just the point sources used in this study.

\begin{figure}[h!]
\includegraphics[scale = 0.55]{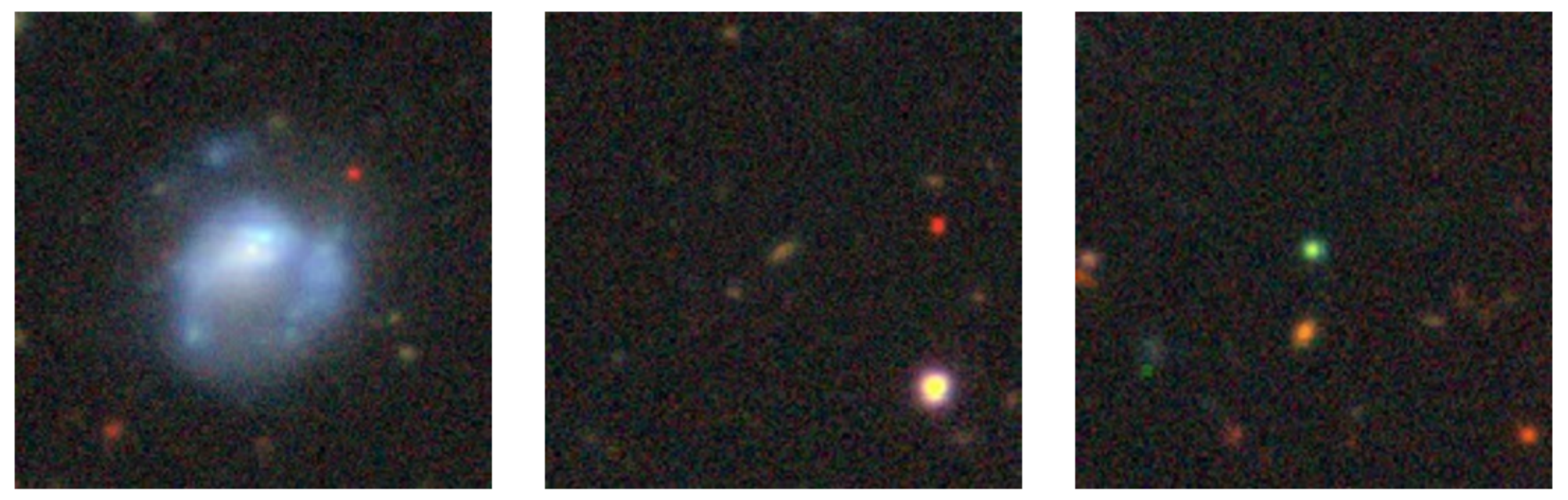}
\centering
\caption{\footnotesize{Visual inspection examples of contamination using image cutouts from DECam for objects classified as quasars in this study. Left: Obvious galaxy contaminant selected by our algorithm. This object is a low-redshift galaxy which has similar $u-g$ colors to quasars with $z \leq 3$. Center: An object selected by our algorithm that exhibits extended behavior, but is not visually an obvious contaminant like the galaxy in the left panel. Objects like this are removed in our final clustering result; spectroscopic follow-up is needed to classify these objects as galaxies. Right: A known high-$z$ quasar that we also select using our algorithm. This particular object is at $z \sim 3.7$ and is a typical point source commonly associated with quasar activity. The co-added color of this quasar appears to be green, however quasars in this study can have a range of colors in DECam Legacy Survey cutouts, depending on their redshift. Each frame is $\sim$ 45$\arcsec$ on a side.}}\label{fig:contampics}
\end{figure}

Figure \ref{fig:contampics} depicts three types of objects that passed the high-$z$ quasar selection algorithms (in either redshift range). In the left panel we show local galaxies ($z\sim 0.3$) which, as a result of the 4000\AA\ break in their spectra, can be mistaken for the Lyman-$\alpha$ forest from high-$z$ quasars (at $z\sim 3.5$). This confusion causes the low-$z$ galaxies to pass the machine-learning selection. These are obvious contaminants that were easily detected and removed by hand. 

We also selected objects that appeared to have extended emission; an example of which is shown in the center panel of Figure \ref{fig:contampics}. While these objects could be galaxies at higher redshift (e.g.,\ Lyman-Break galaxies; \citealt{He2017}), it is also possible that they could be faint quasars at high-$z$ whose emission from the central engine is not bright enough to outshine the host galaxy. For the faint quasars in our study, this could certainly be the case. These objects did not pass our final point source metric and thus were removed from our final analysis.

Finally, in the right-hand panel of Figure \ref{fig:contampics}, we show a known quasar at redshift $z \sim$ 3.7 which our machine-learning algorithm also classifies as a high-$z$ quasar. This object passes the point-source metric and is thus included in this study. Most of the objects that we call point like have similar profiles to this object (albeit, some are much fainter). Once again, spectroscopic follow-up is needed on these objects as well for a combination of testing the classification and testing the redshifts estimates from our machine-learning algorithms. 

\subsection{Error Estimates and Fitting Parameters} \label{append:Errors}

\begin{figure}[h!]
\includegraphics[scale = 0.55]{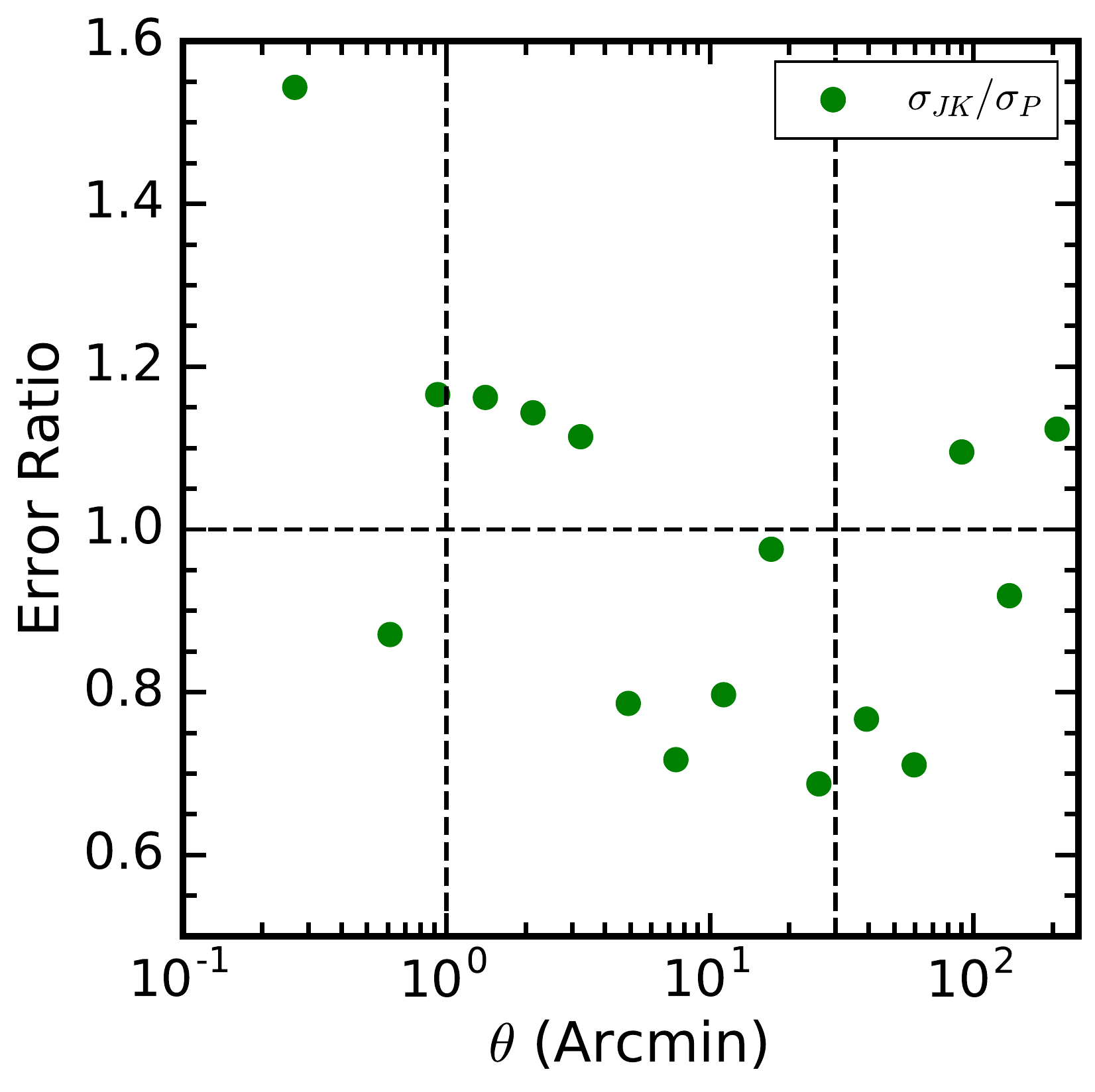}
\centering
\caption{\footnotesize{Ratio of the Jackknife errors to Poisson errors for the full quasar candidate sample. Poisson errors were computed using the pair counts reported in Table \ref{tab:Counts}. In this investigation, we replace the Jackknife errors with Poisson errors wherever the ratio is less than unity.}}\label{fig:JKPratio}
\end{figure}

To ensure that we obtain reasonable jackknife errors, we compare our errors in Table \ref{tab:Counts} to the Poisson errors \citep{Peebles1973} defined as:
\begin{equation}\label{eqn:Poisson}
\sigma_{Poisson} = \frac{1+\omega(\theta)}{\sqrt{DD}}.
\end{equation}
Poisson error is a measure of the noise due to the number of pairs in the sample \citep{Ross2009}, and is most valid at smaller scales where pairs of objects are independent of each other \citep{Shanks1994}. We depict the ratio of our Jackknife errors to the Poisson errors in Figure \ref{fig:JKPratio}. Poisson errors represent a minimum standard deviation in a clustering measurement, particularly on the smallest scales, thus the ratio of the Jackknife to Poisson errors should be of order unity. In this investigation, we replace the Jackknife errors with Poisson errors wherever the ratio of the two in Figure \ref{fig:JKPratio} is less than one.

\begin{figure}[h!]
\includegraphics[scale = 0.45]{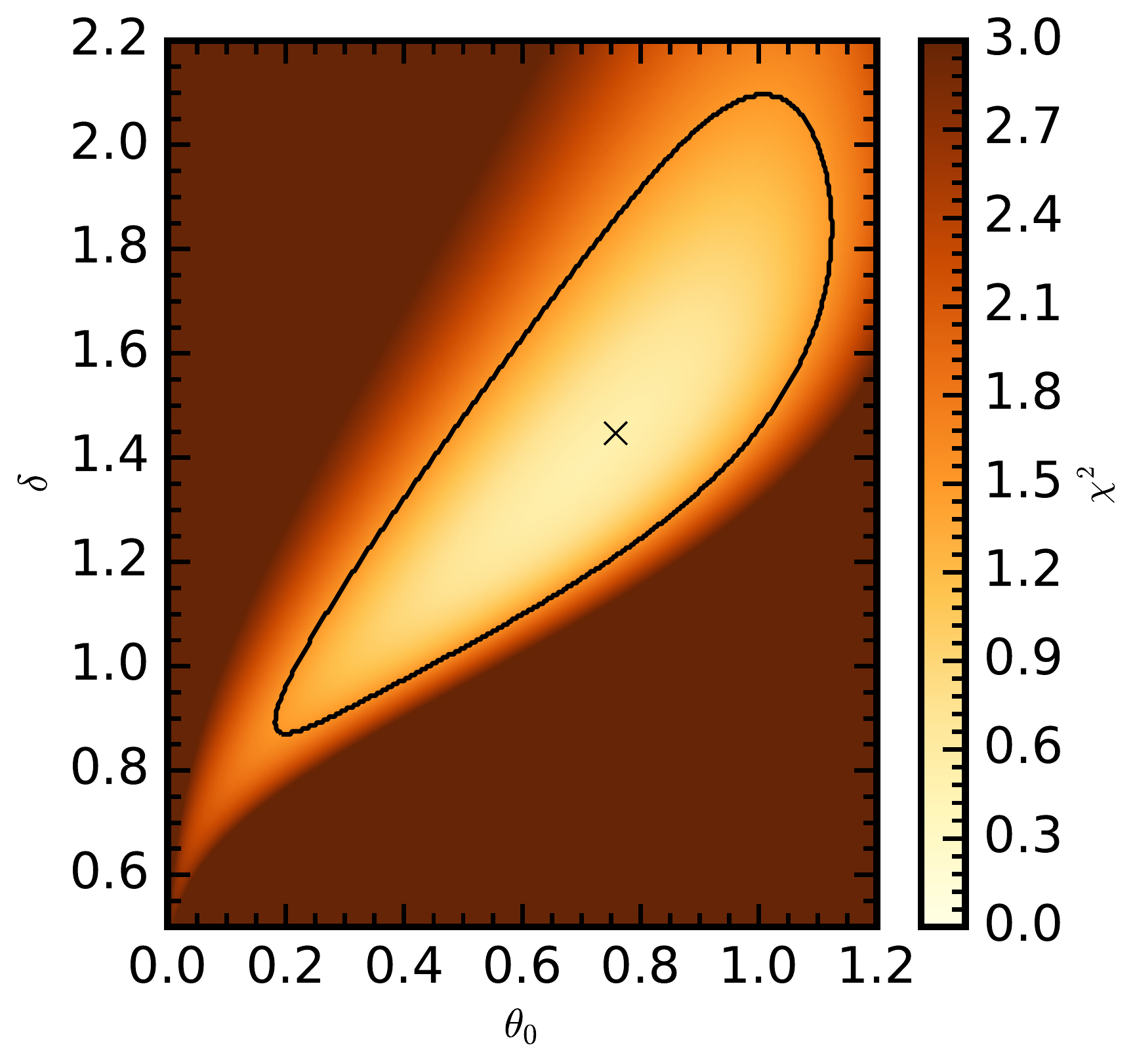}
\includegraphics[scale = 0.45]{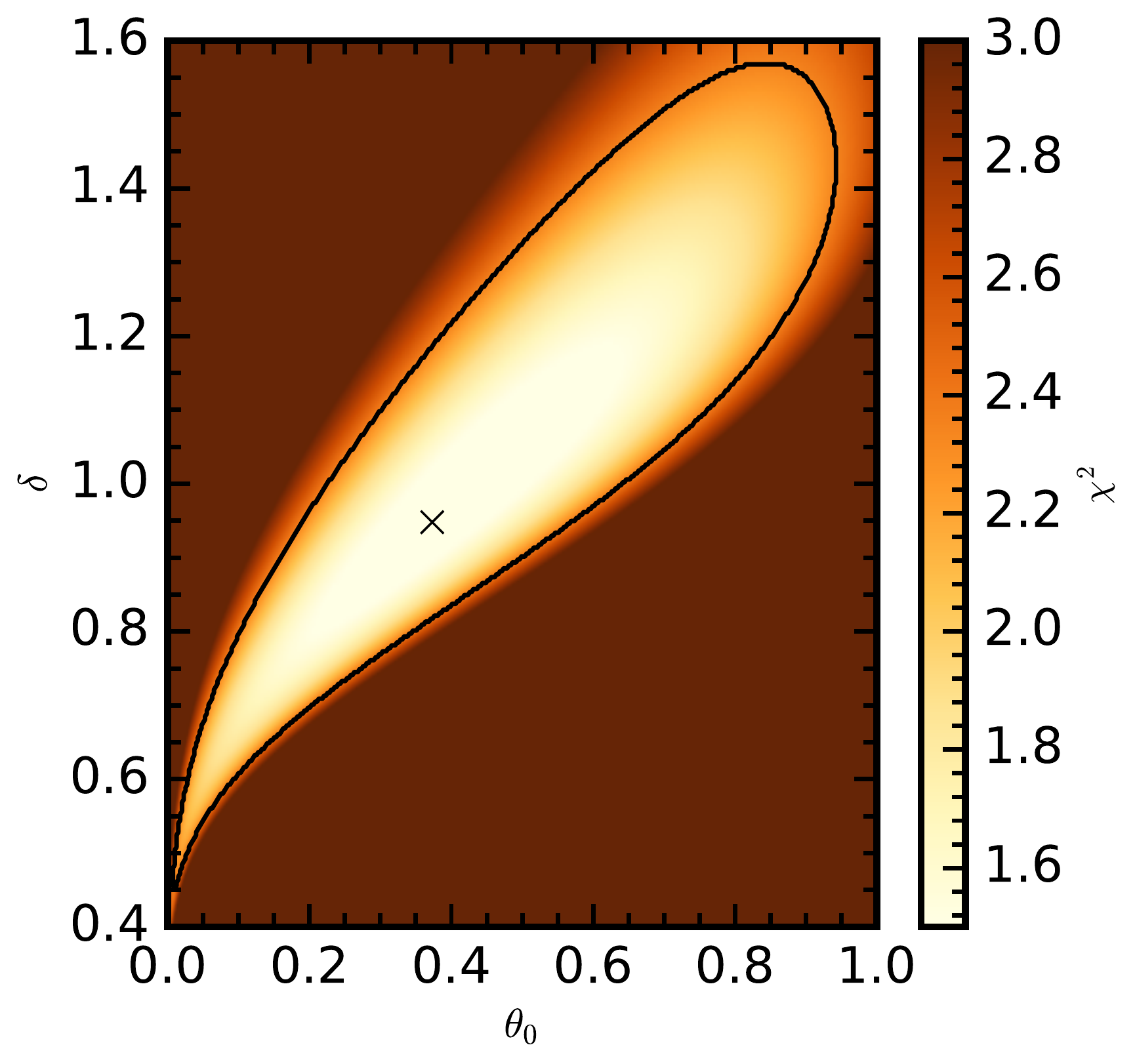}
\centering
\caption{\footnotesize{$\chi^2$ map of the free parameters in the power law model the full (left) and faint (right) samples of photometrically selected quasars. The black point depicts the location of the minimum value of $\chi^2$ which corresponds to the best fit values in Table \ref{tab:bz}. The black contour outlines the 1$\sigma$ region in this space. The range in sigma in both dimensions is representative of the range recorded in Table \ref{tab:bz}.}}\label{fig:plawcov}
\end{figure}

We also test the best fit parameters from both the power law model and the dark-matter model by generating $\chi^2$ maps for each space. We compute the $\chi^2$ as:
\begin{equation}
\chi^2 = \sum \frac{(\omega_{\mathrm{measured}}(\theta) - \omega_{\mathrm{model}}(\theta))^2}{\sigma^2}
\end{equation}
For our power law model, we iterate the power law index over the range $0.5 \leq \delta \leq 2.2$ in 300 steps and the correlation angle $0 \leq \theta_0 \leq 1.2$ in 400 steps and compute the $\chi^2$ value. Figure \ref{fig:plawcov} depicts the results of this analysis for the full sample of quasar candidates (left) and the faint sample (right). In both cases, we find that the best-fit parameters given in Table \ref{tab:bz} (represented by the black `x') lie in the region of the minimum $\chi^2$. We also plot the 1$\sigma$ region and find that it is consistent with the ranges given in Table \ref{tab:bz}.

\begin{figure}[h!]
\includegraphics[scale = 0.45]{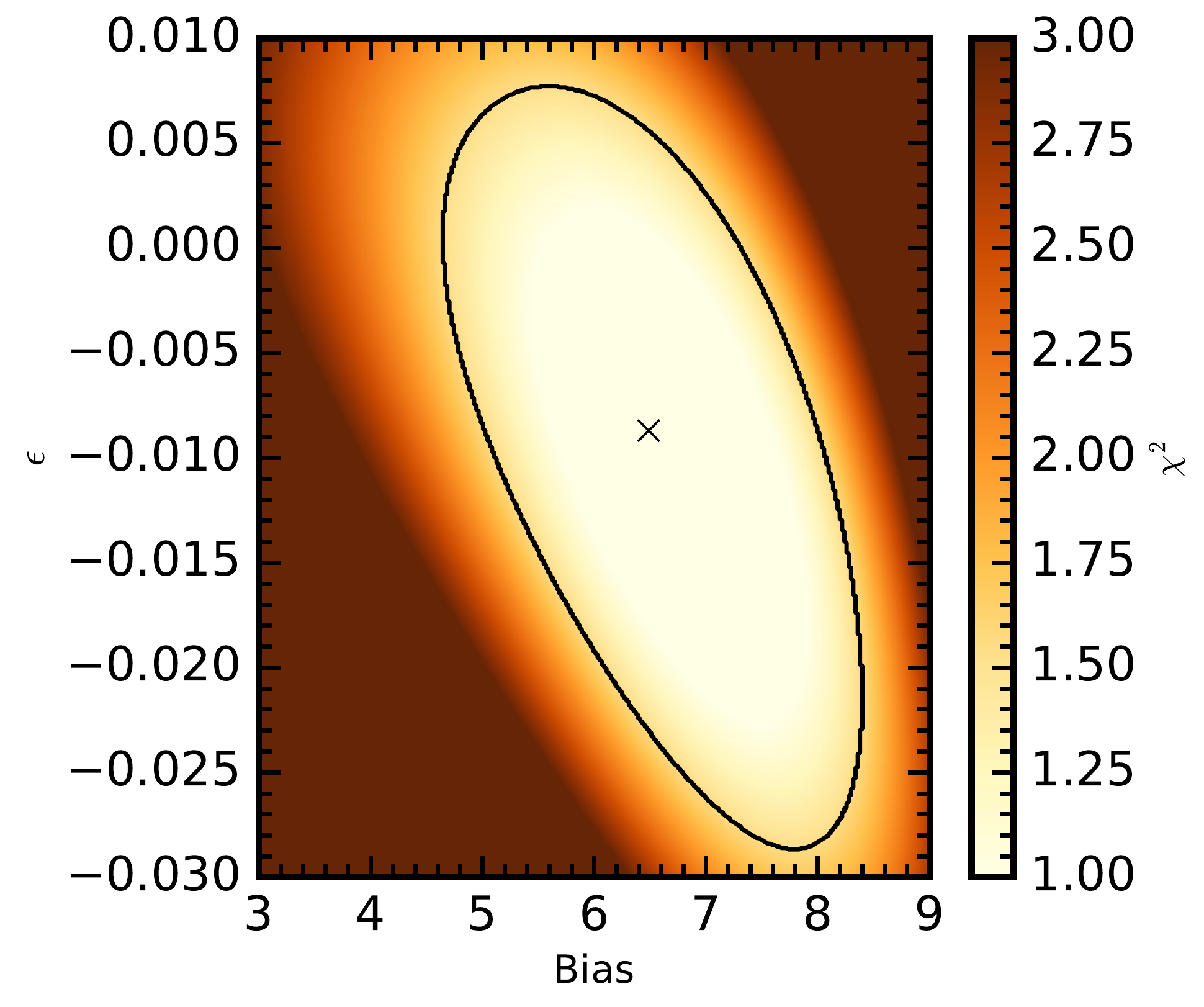}
\includegraphics[scale = 0.45]{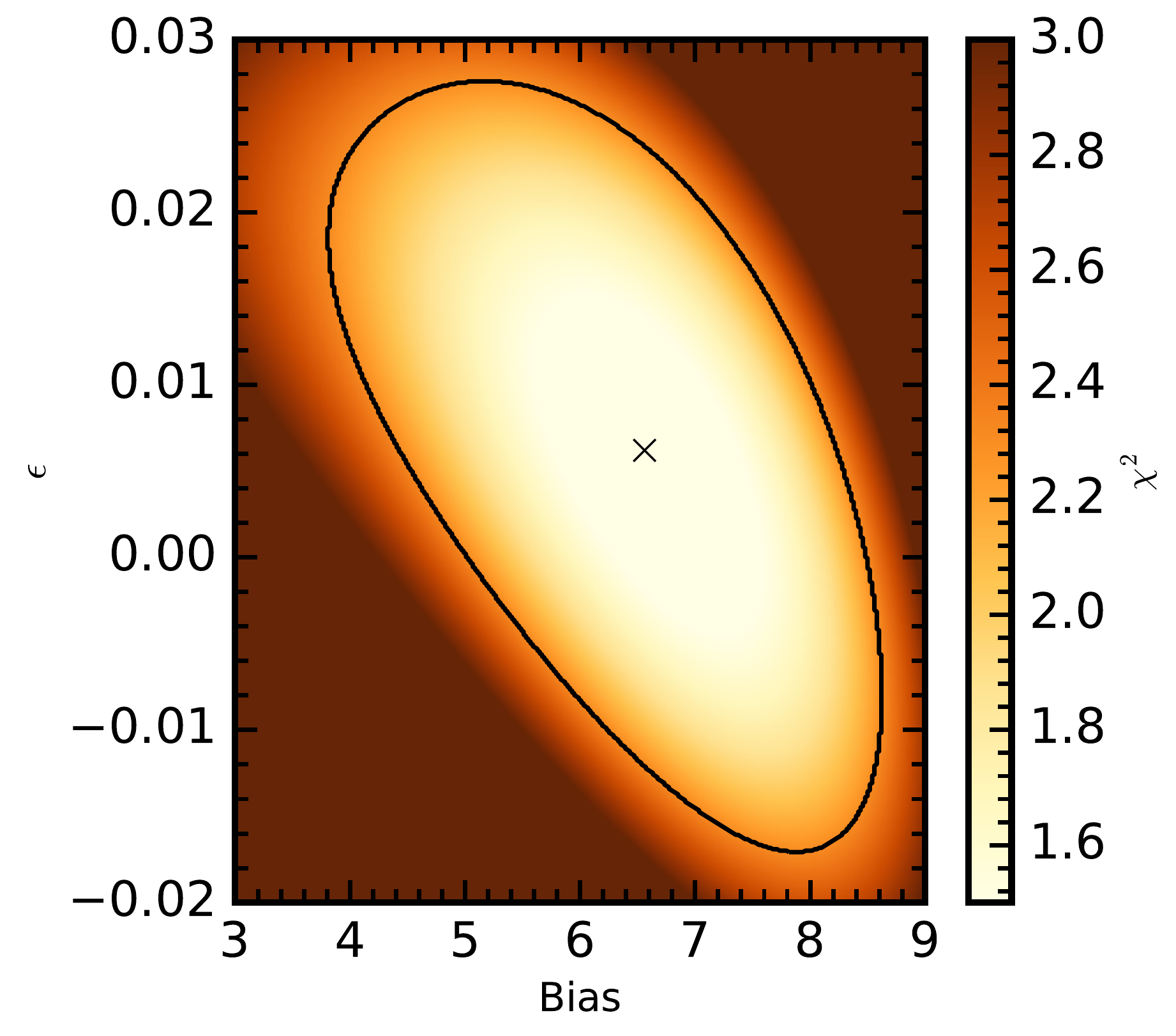}

\centering
\caption{\footnotesize{$\chi^2$ map of the free parameters in the DM model for the full (left) and faint (right) samples of photometrically selected quasars. As in Figure \ref{fig:plawcov}, the black point depicts the location of the minimum $\chi^2$ corresponding to the values in Table \ref{tab:bz}. The black contour outlines the 1$\sigma$ region in this space, and reflects the errors presented in Table \ref{tab:bz}.}}\label{fig:dmmodcov}
\end{figure}

Figure \ref{fig:dmmodcov} depicts the $\chi^2$ map of the dark-matter model, and is computed in a similar manner as before. Here, we iterate both the bias values over a range of $3 \leq \mathrm{b} \leq 9$ in 300 steps, and the cross-correlation term over a range $-0.03 \leq \epsilon \leq 0.01$ in 600 steps. Again, we find that the values reported in Table \ref{tab:bz} are consistent with the minimum $\chi^2$ value, and the errors span an appropriate range.

\newpage
\bibliographystyle{yahapj}
\bibliography{ms}

\end{document}

%% file: tab1.tex
NDWFS+AGES   &          7.9 &           585 & $I\leq$21.5, $[3.6]=6.4\mu$m    & X+R+MIR    & $0.25<z<0.8$   & C/s       & \citet{Hickox2009}\\ 
NDWFS+AGES   &           9  &          924  & $R\sim$25.0, $[3.6]=6.4\mu$m    & opt+MIR        & $0.7<z<1.8 $   & C/b & \citet{Hickox2011}\\ 
PRIMUS+DEEP2 &     $\sim$10 &   $\sim$1 000 & $i_{\rm AB}\sim$23.5            & X+R+MIR     & $0.2<z<1.2$    & C/s       & \citet{Mendez2016}\\ 
{\bf SpIES+SHELA} & {\bf $\approx$100}  & {\bf \hbox{1 378}} & $i\sim$23.5, $[3.6]=6.1\mu$m  & {\bf opt+MIR}   & $\mathbf{z>2.9}$   & {\bf A/b}  & {\bf This study} \\
2SLAQ        & $\approx$150 &  \hbox{ 6 374} & $20.85 < g < 21.85$             & cb/UVX       & $0.3<z<2.9$    & A/s   & \citet{daAngela2008}  \\
HSC          & 172          &  \hbox{ 901} & $21.0 < i < 23.5$                 & opt+NIR       & $3.4<z<4.6$    & C/b   & \citet{He2017}  \\
ACTxSDSS     &          324 &   $\sim$24 000 & $17.75 < i <22.45$              & $^{b}$XDQSO           & $z\approx1.4$  & C/p & \citet{Sherwin2012} \\ 
2QZ          & $\approx$445 &  \hbox{13 989} & $18.25 < b_{\rm J} < 20.85$     & cb/UVX       & $0.8<z<2.1$    & A/s   & \citet{Porciani2004} \\
2QZ          &          721 &  \hbox{22 655} & $18.25 < b_{\rm J} < 20.85$     & cb/UVX       & $0.3<z<2.2$    & A/s   & \citet{Croom2005} \\
eBOSS Y1Q    &         1168 &   $\sim$70 000 & $g \leq 22.0$ or $r \leq 22.0$  & XDQSO                 & $0.9<z<2.1$    & A/s   & \citet{RodriguezTorres2016} \\
eBOSS        &         1200 &  $\sim$69 000  & $g \leq 22.0$ or $r \leq 22.0$  & XDQSOz                & $0.9<z<2.2$    & A/p   & \citet{Laurent2017} \\
eBOSS BAO    &         2044 &  147 000       & $g \leq 22.0$ or $r \leq 22.0$  & XDQSOz                & $0.8<z<2.2$    & A/s   & \citet{Ata2017} \\
SPTxWISE     &         2500 & \hbox{107 469} & $W2\leq15$                      & IR                    & $\left\langle z \right\rangle \sim 1$ & C/p & \citet{Geach2013} \\ 
BOSSxLy$\alpha$ &      3275 &  \hbox{61 342} & $g \leq 22.0$ or $r \leq 21.85$ & XDQSO                 & $2.0<z<3.5$    & C/s  & \citet{FontRibera2013} \\
WISE 	     &         3363 & \hbox{176 467} & $W2<15.05$                      & IR                    & $z\sim1$       & A/p & \citet{Donoso2014} \\
WISE 	     &         3422 & \hbox{175 911} & $W2<15.05$                      & IR                    & $z\sim1$       & A/C/p  & $^{c}$\citet{DiPompeo2016} \\
BOSS DR9     &         3600 &  \hbox{27 129} & $g \leq 22.0$ or $r \leq 21.85$ & XDQSO                 & $2.2<z<2.8 $   & A/s   & \citet{White2012} \\
SDSS DR5     &   $\sim$4000 &  \hbox{38 208} & $i \leq 19.1$                   & cb           & $0.1<z<5.0$    & A/C/s   & \citet{Shen2009} \\
SDSS DR5     &         4013 &  \hbox{30 239} & $i \leq 19.1$                   & cb          & $0.3<z<2.2$    & A/s   & \citet{Ross2009} \\
SDSS DR5     &         4041 &  \hbox{ 4 426} & $i \leq 20.2$                   & cb          & $2.9<z<5.4$    & A/s   & \citet{Shen2007} \\
SDSS DR4     &   $\sim$6670 & $\sim$300 000  & $g < 21$                        & $^{d}$KDE cb & $0.75<z<2.28$  & A/p  & $^{e}$\citet{Myers2007} \\ 
BOSS DR12    &         6950 &  \hbox{55 826} & $g \leq 22.0$ or $r \leq 21.85$ & XDQSO                 & $2.2<z<2.8$    & A/s   & \citet{Eft2015} 

%% file: tab2.tex
0.076 & 0 & 7 & 828 & -0.754 & 0.5038 & -- \\
0.116 & 0 & 17 & 1980 & -0.781 & 0.5074 & -- \\
0.175 & 0 & 37 & 4370 & -0.756 & 0.2472 & -- \\
0.266 & 2 & 116 & 10432 & 0.756 & 2.7091 & 1.7556 \\
0.403 & 2 & 268 & 23350 & -0.459 & 0.8997 & 0.5406 \\
0.611 & 8 & 542 & 53470 & 0.507 & 0.6561 & 0.7534 \\
0.927 & 12 & 1162 & 120784 & 0.073 & 0.5107 & 0.4381 \\
1.405 & 32 & 2652 & 273444 & 0.247 & 0.3623 & 0.3118 \\
2.131 & 74 & 6022 & 619802 & 0.269 & 0.2385 & 0.2086 \\
3.231 & 156 & 13782 & 1403064 & 0.158 & 0.1461 & 0.1312 \\
4.899 & 324 & 30643 & 3191350 & 0.100 & 0.0680 & 0.0865 \\
7.428 & 692 & 69474 & 7209140 & 0.034 & 0.0398 & 0.0556 \\
11.262 & 1506 & 155010 & 16201178 & 0.015 & 0.0295 & 0.0370 \\
17.075 & 3226 & 343452 & 36027696 & -0.014 & 0.0239 & 0.0245 \\
25.889 & 7104 & 747011 & 78725580 & 0.002 & 0.0116 & 0.0168 \\
39.253 & 14932 & 1581774 & 166710784 & -0.005 & 0.0088 & 0.0115 \\
59.516 & 29674 & 3141776 & 330927082 & -0.005 & 0.0058 & 0.0082 \\
90.237 & 53028 & 5579277 & 583004272 & -0.007 & 0.0067 & 0.0061 \\
136.818 & 75010 & 7795784 & 815069184 & 0.006 & 0.0048 & 0.0052 \\
207.443 & 100858 & 10584579 & 1113270342 & 0.002 & 0.0050 & 0.0045